\documentstyle[preprint,aps,eqsecnum,epsf]{revtex}

\begin{document}

\title {\null\vspace*{-.0cm}\hfill {\small nucl-th/0010003} \\ \vskip
0.8cm
The Relativistic N-body Problem\\
in a Separable Two-Body Basis}
\author{Cheuk-Yin Wong}
\address{Physics Division, Oak
Ridge National Laboratory,
Oak Ridge, TN 37831, USA\\
Institute of Physics and Applied Physics, Yonsei University,\\
Seoul 120-749, Korea\\
}
\author{Horace W. Crater}
\address{Department of Physics,
University of Tennessee Space Institute,\\
Tullahoma, TN 37388, USA\\
}
\maketitle

\begin{abstract}
We use Dirac's constraint dynamics to obtain a Hamiltonian formulation
of the relativistic $N$-body problem in a separable two-body basis in
which the particles interact pair-wise through scalar and vector
interactions. The resultant $N$-body Hamiltonian is relativistically
covariant.  It can be easily separated in terms of the center-of-mass
and the relative motion of any two-body subsystem.  It can also be
separated into an unperturbed Hamiltonian with a residual
interaction. In a system of two-body composite particles, the
solutions of the unperturbed Hamiltonian are relativistic two-body
internal states, each of which can be obtained by solving a
relativistic Schr\"odinger-like equation.  The resultant two-body wave
functions can be used as basis states to evaluate reaction matrix
elements in the general $N$-body problem. We prove a relativistic
version of the post-prior equivalence which guarantees a unique
evaluation of the reaction matrix element, independent of the ways of
separating the Hamiltonian into unperturbed and residual
interactions. Since an arbitrary reaction matrix element involves
composite particles in motion, we show explicitly how such matrix
elements can be evaluated in terms of the wave functions of the
composite particles and the relevant Lorentz transformations.
\end{abstract}

\newpage

\section{Introduction}

The theoretical description of the relativistic $N$-body problem is an
interesting but perplexing problem in nuclear and particle physics. It
involves the possibility of relativistic bound states of composite particles
on the one hand and the reaction matrix elements between these particles at
high energies on the other hand. The relativistic $N$-body problem is
complicated by the fact that the relativistic bound state problem is
basically non-perturbative in nature and cannot be solved by the
conventional perturbative quantum field theory. One also needs to describe
the reaction process relativistically by making use of the results from the
bound states.

Much progress has been made in the study of relativistic two-body
bound state problems
\cite{dirac,cnstr,nogo,cra96,cra82,rob,cra92,saz85,tod,cra88,saz89,Bla66,Lep74,Man87,Bij97,Gro69}.
In the 1970s, several authors used Dirac's constraint mechanics
\cite{dirac} to attack the relativistic two-body problem at its
classical roots \cite{cnstr}, successfully evading the so-call ``no
interaction theorem'' \cite{nogo}. The quantum version of the
constraint approach was extended to pairs of spin one-half
particles. \ The results were two-body quantum bound state equations
that correct the defects in the Breit equation, correct the defects in
the ladder approximation to the Bethe-Salpeter equation, and control
covariantly the relative time and energy variables \cite{cra96}. Those
bound state equations for fermions are the two-body Dirac equations of
constraint dynamics, which we shall also call the constraint equations
\cite{cra82}. They possess a number of important desirable features,
some of which are unique. For example, they remove the deficiencies of
earlier approaches in which the spin dependence of the potentials is a
patchwork of semirelativistic corrections determined by field
theory\cite{rob}. In contrast, the spin dependence in constraint
dynamics is determined naturally by the Dirac equation structure of
each of the constraint equations together with the assumed covariant
interactions \cite{cra82,cra92}. The equations of constraint dynamics
are manifestly covariant while yielding simple three-dimensional
Schr\"{o}dinger-type equations, like those of their nonrelativistic
counterparts\cite{cra82,cra92}. (This particular feature will
ultimately be of crucial importance in the discussions given in this
paper.) These constraint equations have passed numerous tests showing
that they reproduce correct QED perturbative results when solved
nonperturbatively\cite{cra92}. In addition, the Dirac forms of these
equations automatically make it unnecessary to introduce {\it ad hoc}
cutoff parameters, which are needed in most other approaches
\cite{rob} to regularize singular potentials. The relativistic
potentials appearing in the constraint equations are related directly
to the interactions of perturbative quantum field theory. In
non-perturbative QCD as applied to meson spectroscopy, they may be
introduced phenomenologically \cite{cra88} and can be regarded as an
anticipation of potentials that may eventually emerge from lattice
gauge theory.

In the present paper we will extend the constraint equations beyond
the scope of the two-body problem and will, for simplicity, limit
ourselves to the constraint description of spinless particles. We
first summarize the constraint approach for two spinless
particles. For each particle, one ascribes a generalized mass shell
constraint which includes the interaction.  The constraints must be
consistent with each other and this in turn restricts the dependence
of the interactions on the relative coordinates, eliminating both the
relative time and energy variables in the CM system.  The resultant
equations correspond to a Bethe-Salpeter equation whose kernel and
Green's functions are constrained by the requirement of $P\cdot q$ =0,
where $P$ and $q$ are the total and relative momenta of the two-body
system \cite{saz85}. In particular, from the Bethe-Salpeter equation
with this constraint, one can derive \cite{cra88,cra92} the
``quasipotential equation'' of Todorov\cite{tod}, which is a
Schr\"{o}dinger-like inhomogeneous integral equation where the
quasipotential $\Phi $ is related to the scattering amplitude in
perturbative quantum field theory.  Other methods of reducing the
Bethe-Salpeter equation have also been suggested
\cite{Bla66,Lep74,Man87,Bij97,Gro69}.

At present, the relativistic treatment of the $N$-body problem and the
reaction of composite particles at high energies have not advanced as
much as in the relativistic treatment of two-body bound states. An
investigation of the relativistic $N$-body problem was previously
carried out by Sazdjian \cite{saz89}. In contrast, the
non-relativistic $N$-body problem and the non-relativistic description
of the reaction of composite particles have been well developed
\cite{Sat80,Bar92,Bro72,Rin80}. One has, for example, the
distorted-wave Born approximation method in nuclear reactions \cite
{Sat80} and the quark-interchange model in hadron reactions
\cite{Bar92} for reactions between composite particles. One first solves
for the wave functions and determines the interaction between the
constituents using the energy levels of bound states. Then one uses
the same set of interactions and wave functions to calculate the
reaction matrix elements for the reaction of composite
particles. There is also the potentially different values of the post
or prior forms of the reaction matrix element, which distinguish
whether the interaction occurs before the rearrangement or after the
rearrangement of the constituents. The post-prior equality of the
reaction matrix elements is attained when the interaction and wave
functions that are used in the calculation of the overlap integral are
the same as the interaction and the wave functions obtained in the
bound state analysis.

The formalism we shall develop can be applied to many processes of interest.
For example, in high-energy heavy-ion collisions, the investigation of the
dynamics and the properties of the produced hadron matter involve the
reaction cross sections and the reaction matrix elements between the produced
hadrons at relativistic energies. Most of these cross sections and reaction
matrix elements cannot be measured experimentally. A reliable theoretical
relativistic model which describes the $N$-body problem of the constituents
is needed for their evaluation. It is necessary to generalize the
nonrelativistic reaction model to study reactions between composite
particles at relativistic energies. It is also of interest to study the
relativistic $N$-body problem in order to investigate an assembly of
composite particles and their clustering or molecular states.

We shall take advantage of previous advances in the understanding of
the relativistic two-body problem in our present study of the
relativistic $N$-body problem. In this work, we specialize in $N$-body
dynamics with just pair-wise interactions between particles. We shall
write down the Hamiltonian formulation of the relativistic $N$-body
problem which allows an easy separation in terms of the center-of-mass
and the relative motion for any two-body system and a simple
separation of the unperturbed Hamiltonian and the residual
interaction. We can then use the relativistic two-body bound states as
basis states for the investigation of the relativistic $N$ -body
problem. We can construct a proof of the
``post-prior'' equivalence in relativistic dynamics, which guarantees
that the reaction matrix element is independent of the different ways
of partitioning the unperturbed Hamiltonian and the residual
interaction from the $N-$body Hamiltonian.

This paper is organized as follows. In Section II we review important
aspects of the constraint approach and introduce the relativistic
two-body Hamiltonian and bound state eigenvalue equation. We show that
the relativistic two-body solution in the CM rest system has the
simplicity of its nonrelativistic counterparts. In Section III we
generalize the two-body Hamiltonian to the case of $N$ particles. We
discuss some applications of the present formulation in Section IV. We
show how to utilize the constituent two-body wave functions for the
$N$-body problem. Because of the analogy to their nonrelativistic
counterparts, these states can be used as basis states to evaluate a
general reaction matrix element in the general $N$-body problem. We
prove the relativistic version of the post-prior equivalence for the
reaction matrix elements. As an illustration and a problem of
practical interest, we consider in Section V the reaction of four
particles in two composite systems where the non-relativistic
treatment has already been formulated by Barnes and Swanson
\cite{Bar92}. In Section VI, we explicitly obtain the wave function in
second-quantized form so as to construct the overlap integral for the
reaction matrix element. Such a reaction matrix element 
involves states of composite
particles in motion. Thus we show in Section VII explicitly how to
evaluate such elements in terms of the wave functions of each
composite particle, and develop the relevant Lorentz transformation
laws required. \ Section VIII summarizes and points to future
problems.

\section{Hamiltonian formulation of the 2-Body Problem from Constraint
Dynamics}

\bigskip

We can formulate the relativistic treatment of the two-body problem for
spinless particles \cite{Cra84,cra81} in a way that has the simplicity of
the ordinary non-relativistic two-body Schr\"{o}dinger equation and yet
maintains relativistic covariance. Including spin and generalizing to
different types of interactions can be carried out in a more complete
framework \cite{cra90,cra96}.

For each particle we assume a generalized mass shell constraint of the form 
\begin{equation}
{\cal H}_{i}|\psi \rangle=0{~~~~~~{\rm for~~~~~~~~~}}i=1,2  \label{1}
\end{equation}
where 
\begin{equation}
{\cal H}_{i}=p_{i}^{2}-m_{i}^{2}-\Phi _{i},
\end{equation}
and $\Phi _{1}$ and $\Phi _{2}$ are two-body interactions dependent on $%
x_{12}$. One constructs the total Hamiltonian ${\cal H}$ from these
constraints by 
\begin{equation}
{\cal H=\lambda }_{1}{\cal H}_{1}+\lambda _{2}{\cal H}_{2},  \label{2.5}
\end{equation}
(with $\lambda _{i}$ as Lagrange multipliers). \ In order that each of these
constraints be conserved in time we must have 
\begin{equation}
\lbrack {\cal H}_{i},{\cal H}]|\psi \rangle=i\frac{d{\cal H}_{i}}{d\tau }%
|\psi \rangle=0.
\end{equation}
As a consequence, we have 
\begin{eqnarray}
\lbrack {\cal H}_{i},{\cal \lambda }_{1}{\cal H}_{1}+\lambda _{2}{\cal H}%
_{2}]|\psi &\rangle&=  \nonumber \\
\{[{\cal H}_{i},\lambda _{1}]{\cal H}_{1}|\psi &\rangle&+\lambda _{1}[{\cal H%
}_{i},{\cal H}_{1}]|\psi \rangle+[{\cal H}_{i},\lambda _{2}]{\cal H}%
_{2}|\psi \rangle+\lambda _{2}[{\cal H}_{i},{\cal H}_{2}]\}|\psi \rangle=0.
\end{eqnarray}
\ Using Eq.(\ref{1}), the above equation leads to the compatibility
condition between the two constraints, 
\begin{equation}  \label{eq:com}
\lbrack {\cal H}_{1},{\cal H}_{2}]|\psi \rangle=0.  \label{3}
\end{equation}
Since the mass commutes with the operators, this implies 
\begin{equation}
{\large (}[p_{1}^{2},\Phi _{2}]+[\Phi _{1},p_{2}^{2}]+[\Phi _{1},\Phi _{2}]%
{\large )}|\psi \rangle=0.  \label{7}
\end{equation}
The simplest way to satisfy the above equation is to take 
\begin{equation}  \label{eq:res}
\Phi _{1}=\Phi _{2}=\Phi (x_{\perp }),  \label{8}
\end{equation}
which is a kind of relativistic Newton's third law. Here, the transverse
coordinate is defined by 
\begin{equation}
x_{\nu \perp }=x_{12}^{\mu }(\eta _{\mu \nu }-P_{\mu }P_{\nu }/P^{2}),
\label{9}
\end{equation}
where $P$ is the total momentum 
\begin{equation}
P=p_{1}+p_{2}.
\end{equation}
The choice of the two-body potential Eq. (\ref{eq:res}) leads to 
\begin{equation}
\lbrack {\cal H}_{1},{\cal H}_{2}]|\psi \rangle=2P\cdot \partial_{x_{12}} \Phi
(x_{\perp })|\psi \rangle=0,
\end{equation}
and the compatibility condition (\ref{eq:com}) is satisfied.

The two-body Hamiltonian ${\cal H}$ determines the dynamics of the two-body
system. Its equation of motion is 
\begin{equation}
{\cal H}|\psi \rangle=0.
\end{equation}

This equation describes both the center-of-mass motion and the internal
relative motion. To characterize the center-of-mass motion, we note that
since the potential $\Phi $ depends only on the difference of the two
coordinates we have 
\begin{equation}
\lbrack P,{\cal H}]|\psi \rangle=0.
\end{equation}
(This does not require that $[P,\lambda _{i}]=0$ since the ${\cal H}%
_{i}|\psi \rangle=0$.) Thus, $P$ is a constant of motion and we can take $%
|\psi \rangle$ to be an eigenstate state characterized by a total momentum $%
P $.

To separate out the internal relative motion from the center-of-mass
motion, we introduce the relative momentum $q$ defined by
\begin{equation}
p_{1}={\frac{p_{1}\cdot P}{P^{2}}}P+q,  \label{eq:213}
\end{equation}
\begin{equation}
p_{2}={\frac{p_{2}\cdot P}{P^{2}}}P-q,  \label{eq:214}
\end{equation}
where the first term on the right hand side of the above two equations is
the projection of each momentum onto the total momentum. The above
definition of the relative momentum guarantees the orthogonality of the
total momentum and the relative momentum,
\begin{equation}
P\cdot q=0,  \label{pq}
\end{equation}
which follows from taking the scalar product of either equation with $P$.
From Eqs.\ (\ref{eq:213}) and (\ref{eq:214}) this relative momentum can be
written in terms of $p_{1}$ and $p_{2}$ as 
\begin{equation}
q=\frac{\varepsilon _{2}}{\sqrt{P^{2}}}p_{1}-\frac{\varepsilon _{1}}{\sqrt{%
P^{2}}}p_{2}  \label{qrel}
\end{equation}
where 
\begin{eqnarray}
\varepsilon _{1} &=&{\frac{p_{1}\cdot P}{\sqrt{P^{2}}}}=\frac{%
P^{2}+p_{1}^{2}-p_{2}^{2}}{2\sqrt{P^{2}}}  \nonumber \\
\varepsilon _{2} &=&{\frac{p_{2}\cdot P}{\sqrt{P^{2}}}}=\frac{%
P^{2}+p_{2}^{2}-p_{1}^{2}}{2\sqrt{P^{2}}}  \label{ep12}
\end{eqnarray}
are the projections of the momenta $p_1$ and $p_2$ along the direction
of the total momentum $P$.  Using Eqs. (\ref{1}) and (\ref{8}) and
taking the difference of the two constraints, we obtain
\begin{equation}
(p_{1}^{2}-p_{2}^{2})|\psi \rangle=(m_{1}^{2}-m_{2}^{2})|\psi \rangle.
\end{equation}
Thus on these states $|\psi \rangle$ we have 
\begin{eqnarray}
\varepsilon _{1} &=&\frac{P^{2}+m_{1}^{2}-m_{2}^{2}}{2\sqrt{P^{2}}} 
\nonumber \\
\varepsilon _{2} &=&\frac{P^{2}+m_{2}^{2}-m_{1}^{2}}{2\sqrt{P^{2}}}.
\label{eps}
\end{eqnarray}
Using Eqs.\ (\ref{eq:213}), (\ref{eq:214}), and Eq.(\ref{pq}), we can write $%
{\cal H}$ in terms of $P$ and $q$: 
\begin{eqnarray}
{\cal H}|\psi &\rangle&=\{\lambda _{1}[\varepsilon
_{1}^{2}-m_{1}^{2}+q^{2}-\Phi (x_{\perp })]+\lambda _{2}[\varepsilon
_{2}^{2}-m_{2}^{2}+q^{2}-\Phi (x_{\perp })]\}|\psi \rangle  \nonumber \\
&=&(\lambda _{1}+\lambda _{2})[b^{2}(P^{2};m_{1}^{2},m_{2}^{2})+q^{2}-\Phi
(x_{\perp })]|\psi \rangle=0,  \label{eq:sep}
\end{eqnarray}
where 
\begin{equation}
b^{2}(P^{2},m_{1}^{2},m_{2}^{2})=\varepsilon _{1}^{2}-m_{1}^{2}=\varepsilon
_{2}^{2}-m_{2}^{2}\ =\frac{1}{4P^{2}}%
(P^{4}-2P^{2}(m_{1}^{2}+m_{2}^{2})+(m_{1}^{2}-m_{2}^{2})^{2}).
\end{equation}

Equation (\ref{eq:sep}) contains both the center-of-mass momentum $P$
[through the $b^{2}(P^{2},m_{1}^{2},m_{2}^{2})$] and the relative
momentum $ q $. This constraint equation of $P$ and $q$ can\ then be
solved by the method of the separation of variables. \ That is, we
introduce the bound state eigenvalue $M$ to separate Eq. (\ref{eq:sep})
into the following two equations for the center-of-mass motion and the
internal motion
\begin{equation}
\left\{ P^{2}-M^{2}\right\} |\psi \rangle=0,  \label{pm}
\end{equation}
and 
\begin{equation}
(\lambda _{1}+\lambda _{2})\left\{ q^{2}-\Phi (x_{\perp
})+b^{2}(M^{2},m_{1}^{2},m_{2}^{2})\right\} |\psi \rangle=0,  \label{eig}
\end{equation}
where we have used the first equation on the eigenstate $|\psi
\rangle$ so that $b^{2}(P^{2},m_{1}^{2},m_{2}^{2})$ becomes the
standard triangle function indicative of the presence of exact
relativistic two-body kinematics:
\begin{equation}
b^{2}(M^{2},m_{1}^{2},m_{2}^{2})=\frac{1}{4M^{2}}\left\{
M^{4}-2M^{2}(m_{1}^{2}+m_{2}^{2})+(m_{1}^{2}-m_{2}^{2})^{2}\right\}.
\end{equation}
The eigenvalue equation Eq.\ (\ref{eig}) for the relative motion is
independent of the Lagrange multipliers. It is nonetheless convenient to
choose $\lambda _{i}=1/2m_{i}$ so that the resultant Schr\"{o}dinger
equation matches the non-relativistic two-body Schr\"{o}dinger equation term
by term. Such a choice also helps us obtain useful simplifications in the
relativistic $N$-body problem in later sections. In particular, the $N$-body
Hamiltonian can be easily separated into pairs of two-body Hamiltonians.
This separation makes it easy to introduce the unperturbed Hamiltonian and
residual interactions.

We note that because of the orthogonality of $P$ and $q$, we can write Eq.\ (%
\ref{eig}) in the form 
\begin{equation}
\left( {\frac{1}{2m_{1}}}+{\frac{1}{2m_{2}}}\right) \{q_{\perp }^{2}-\Phi
(x_{\perp })+b^{2}(M^{2},m_{1}^{2},m_{2}^{2})\}|\psi \rangle=0,
\end{equation}
where $q_{\perp }=q-q\cdot PP/P^{2}=q$. [Note that if the relative
momentum were defined in terms of Eq.\ (\ref{eps}) instead of Eq.\
(\ref{ep12}) then we would have
\begin{equation}
q\cdot P|\psi \rangle=0  \label{eqpq}
\end{equation}
but not $q\cdot P=0$ so that $q^{2}|\psi \rangle=q_{\perp }^{2}|\psi
\rangle$.  In either case the coefficients $\varepsilon _{i}$ are
invariant and hence Eq.\ (\ref{qrel}) the same form regardless of
which frame it is evaluated in.]

We show below how the eigenvalue $M$ is related to the eigenvalue obtained
in a non-relativistic Schr\"{o}dinger equation. We go to the
center-of-momentum system where $q=q_{\perp }(0,{\bf q)}$ and $x_{\perp }=(0,
{\bf r)}$ (relative energy and time thus being removed from the problem). We
then have the equation for the relative motion, 
\begin{equation}
\left\{ {\frac{{\bf q}^{2}}{2\mu }}+{\frac{\Phi ({\bf r})}{2\mu }}-{\frac{%
b^{2}}{2\mu }}\right\} |\psi \rangle=0,  \label{eq:eig}
\end{equation}
where $\mu $ is the non-relativistic reduced mass, 
\begin{equation}
\mu ={\frac{m_{1}m_{2}}{m_{1}+m_{2}}}.
\end{equation}
We can cast Eq.\ (\ref{eq:eig}) into the usual form of a non-relativistic
Schr\"{o}dinger equation. By renaming $\Phi /2\mu $ as $V_{12}$, and $%
b^{2}/2\mu $ as $E$, Eq.\ (\ref{eig}) becomes 
\begin{equation}
\left( \frac{{\bf q}^{2}}{2\mu }+V_{12}\right) |\psi \rangle=E|\psi \rangle.
\label{eq:nonrel}
\end{equation}
The above Schr\"{o}dinger equation can be solved to give the eigenvalue $E$.
Then, from the equation $b^{2}(M^{2},m_{1}^{2},m_{2}^{2})=2\mu E$, one can
solve for $M$ in terms of $E$ and obtain 
\begin{equation}
M=\sqrt{2\mu E+m_{1}^{2}}+\sqrt{2\mu E+m_{2}^{2}}.  \label{eq:ME}
\end{equation}
It is easy to show from this that in the limit of very weak binding, the
nonrelativistic limit, we have the familiar result 
\begin{equation}
M=m_{1}+m_{2}+E.
\end{equation}

If one is only interested in the effect of \ exact two-body relativistic
kinematics with $V_{12}$ an energy-independent nonrelativistic potential,
the bound state eigenvalue $M$ for the relativistic two-body problem is
related to the eigenvalue $E$ of the nonrelativistic problem by Eq.\ (\ref
{eq:ME}). It is important to note, however, that the potential $V_{12}$ in
relativistic constraint dynamics includes relativistic dynamical corrections
as well. These corrections include dependences of the potential on the CM
energy $M$ and on the nature of the interaction. For spinless particles
interacting by way of a world scalar interaction $S$, one finds \cite
{cra81,crayng} 
\begin{equation}
V_{12}={\frac{\Phi }{2\mu }}={\frac{2m_{M}S+S^{2}}{2\mu }}  \label{scl}
\end{equation}
where 
\begin{equation}
m_{M}=\frac{m_{1}m_{2}}{M},
\end{equation}
while for (time-like) vector interaction $A$, one finds
\cite{tod,cra81,crayng}
\begin{equation}
V_{12}={\frac{\Phi }{2\mu }}={\frac{2\varepsilon _{M}A-A^{2}}{2\mu }},
\end{equation}
where 
\begin{equation}
\varepsilon _{M}=\frac{M^{2}-m_{1}^{2}-m_{2}^{2}}{2M}  \label{vct}
\end{equation}
and for combined space-like and time-like vector interactions (that
reproduce the correct energy spectrum for scalar QED \cite{Cra84})
\begin{equation}
V_{12}={\frac{\Phi }{2\mu }}={\frac{2\varepsilon _{M}A-A^{2}+\vec{\nabla}
^{2}\log (1-2A/M)^{1/2}+[\vec{\nabla}\log (1-2A/M)^{1/2}]^{2}}{2\mu }}.
\end{equation}
The variables $m_{M}$ and $\varepsilon _{M}$ \ (which both approach $\mu $
in the nonrelativistic limit) were introduced by Todorov \cite{tod} in his
quasipotential approach and are called the relativistic reduced mass and
energy of the fictitious particle of relative motion. \ In the
nonrelativistic limit, $\Phi $ approaches $2\mu (S+A).$ In the relativistic
case, the dynamical corrections to $V_{12}$ referred to above include both
quadratic additions to $S$ and $A$ as well as CM energy dependence through $%
m_{M}$ and $\varepsilon _{M}.$\ This latter point implies that the effective
potential $V_{12}$ depends on the eigenvalue $E$ (or $M$) to be evaluated.
One can obtain the mass of the bound state $M$ by an iterative procedure.
One starts with an estimated $M$ (or $E$) value and obtains the potential $%
V_{12}$. Equations (\ref{eq:nonrel}) and (\ref{eq:ME}) can then be used
iteratively to obtain successively improved values of $V_{12}$ and the
eigenvalue $M$ (or $E$).

We note in passing that since 
\begin{equation}
b^{2}=\varepsilon _{M}^{2}-m_{M}^{2},
\end{equation}
we can write the Schr\"{o}dinger-like equation for combined scalar and
(time-like) vector interactions as \cite{cra81} 
\begin{equation}
\left\{ {\bf q}^{2}+(m_{M}+S)^{2}-(\varepsilon _{M}-A)^{2}\right\} |\psi
\rangle=0,  \label{kgeq}
\end{equation}
which is suggestive of a Klein-Gordon equation for an effective
particle of relative motion. This bound state equation incorporates
not only the correct relativistic kinematics but also the correct
relativistic dynamical corrections through order $1/c^{2}$ and higher,
depending on the input. \ It does it without the necessity of
introducing complicated momentum-dependent Darwin-like interactions,
thereby retaining the simplicity of the non-relativistic
Schr\"{o}dinger equation. Furthermore, the potentials in these
equations are connected to those of Wheeler-Feynman electrodynamics
(and its scalar counterpart) \cite{cra87,cra94}. They have been
obtained systematically from perturbative quantum field theory
\cite{cra92,saz85,cra88} and from an eikonal summation of Feynman
diagrams \cite{sazd}.

In summary, Eqs. (\ref{eq:eig}), (\ref{eq:nonrel}), and (\ref{eq:ME})
provide a useful way to obtain the solution of the relativistic
two-body problem for spinless particles in scalar and vector
interactions. In other works they have been extended to include spin
and have been found to give an excellent account of the bound state
spectrum of both light and heavy mesons using reasonable input quark
potentials \cite{cra88,cra94}.

\section{Hamiltonian formulation of the $N$-Body Problem from Constraint
Dynamics: Separable Two-Body Basis}

The above treatment of the two-body problem can, to some extent, be
generalized to the case of the $N$-body problem \cite{saz89}. \ Our
approach differs and extends the work of Sazdjian in that we choose to
formulate the $N$-body problem in a separable two-body basis.

We consider a system of $N$ particles. For each particle, we specify a
generalized mass shell constraint of the form 
\begin{equation}
{\cal H}_{i}|\psi \rangle=0{~~~~~~{\rm for~~~~~}}i=1,..,N  \label{III1}
\end{equation}
where 
\begin{equation}
{\cal H}_{i}=p_{i}^{2}-m_{i}^{2}-\sum_{j,j\neq i}^{N}\Phi _{ij}-W_{i}.
\end{equation}
The $\Phi _{ij}$ are two-body interactions dependent on $x_{ij}$, and $W_{i}$
are possible $N$-body forces ($N>2)$. We construct the total Hamiltonian
for the system as 
\begin{equation}
{\cal H=}\sum_{i=1}^{N}\lambda _{i}{\cal H}_{i}  \label{4}
\end{equation}
where $\lambda _{i}^{\prime }$s are the Lagrange multipliers. For each
constraint to be conserved in time, we must have 
\begin{equation}
\lbrack {\cal H}_{i},{\cal H}]|\psi \rangle=i(\frac{d}{d\tau }{\cal H}%
_{i})|\psi \rangle=0.
\end{equation}
From Eq.\ (\ref{4}) and Eq.\ (\ref{III1}), we must have the compatibility
condition 
\begin{equation}
\lbrack {\cal H}_{i},{\cal H}_{j}]|\psi \rangle=0.
\end{equation}
Now we attempt to expand out the above equation. For a fixed pair of $i$ and 
$j$, we have 
\begin{eqnarray}
&&{\large (-}[p_{i}^{2},\sum_{k,k\neq j}^{N}\Phi _{jk}]-[\sum_{k,k\neq
i}^{N}\Phi _{ik},p_{j}^{2}]+[\sum_{l,l\neq i}^{N}\Phi
_{il}+W_{i},\sum_{k,k\neq j}^{N}\Phi _{jk}+W_{j}]  \nonumber \\
&&-[p_{i}^{2},W_{j}]-[W_{i},p_{j}^{2}]{\large )}|\psi \rangle  \nonumber \\
&=&0.  \label{6}
\end{eqnarray}
\ Motivated by the form of our two-body solution, we assume that 
\begin{eqnarray}
\Phi _{ij}=\Phi _{ji}=\Phi _{ij}(x_{ij\perp }) 
\end{eqnarray}
in which 
\begin{equation}
(x_{ij\perp })_{\nu }=x_{ij}^{\mu }\left[ \eta _{\mu \nu }-(P_{ij})_{\mu
}(P_{ij})_{\nu }/P_{ij}^{2}\right] ,
\end{equation}
and $P_{ij}=p_{i}+p_{j}$. \ This implies that the $N$-body forces must be
present and satisfy 
\begin{equation}
{\large (}[\sum_{l~l\neq i}^{N}\Phi _{il}+W_{i},\sum_{k\,k\neq j}^{N}\Phi
_{jk}+W_{j}]-[p_{i}^{2},W_{j}]-[W_{i},p_{j}^{2}]{\large )}|\psi \rangle=0.
\end{equation}
These are very complicated equations, and, unlike the two-body case,
there are no known closed-formed solutions \cite{saz89,saz81}. 
Evidently these forces are dependent on the two-body forces themselves
\cite{coes}. In practice, one often ignores these many-body forces and
considers only pair-wise interactions, as we will do in our subsequent
computations of reaction matrix elements in Section VII. That is, under the
approximation in which we set $W_{i}=0$, we can view the particles as
interacting with each other via two-body interactions in a pair-wise
manner. \ However, in most of our formal analysis in this paper, we
will retain these many-body interactions.

The conservation of the constraints in time depends only upon the
compatibility of these constraints and does not depend on the choice
of the Lagrange multipliers. This arbitrariness in the choice of the
$\lambda _{i} $ is similar to a kind of gauge invariance. \ Choosing a
particular set of $\lambda _{i}$ is analogous to choosing a gauge
and can be done for convenience. In the general $N-$body formalism we
will find it convenient to choose $\lambda _{i}=1/2m_{i}$. Such a
choice has many advantages. First it leads to a simple correspondence
with the non-relativistic two-body and $N$-body
Hamiltonians. Secondly this choice depends only on the particle in
question and not what other particle it is linked with. In the general
$N$-body formalism we will find it convenient not to have a
preferred pairing of 2-body composite subsystems. The choice $\lambda
_{i}=m_{i}/2$ avoids this.  Finally, this allows the $N$-body
Hamiltonian to be conveniently separated into a nonperturbative part
and the residual interactions part. Such a correspondence helps one
generalize the post-prior equivalence of reaction matrix elements from
the non-relativistic case to the relativistic case.

With this choice, the relativistic $N$-body Hamiltonian is 
\begin{eqnarray}
{\cal H} &=&\sum_{i=1}^{N}\frac{1}{2m_{i}}(p_{i}^{2}-m_{i}^{2})-
\sum_{i=1}^{N}\sum_{j~j>i}^{N}\frac{\Phi _{ij}}{2\mu _{ij}}-\sum_{i}\frac{
W_{i}}{2m_{i}}  \nonumber \\
&=&\sum_{i=1}^{N}\frac{1}{2m_{i}}(p_{i}^{2}-m_{i}^{2})-\sum_{i=1}^{N}
\sum_{j~j>i}^{N}V_{ij}-\sum_{i}\frac{W_{i}}{2m_{i}},  \label{11}
\end{eqnarray}
where $\mu _{ij}=m_{i}m_{j}/(m_{i}+m_{j})$, and we have introduced the
simplified notation $V_{ij}=\Phi _{ij}/2\mu _{ij}$. (Note that in light of
the above forms of Eqs.\ (\ref{scl}) and (\ref{vct}) for $\Phi _{i}$, the
choice $\lambda _{i}=1/2m_{i}$ \ gives the correct nonrelativistic limit for 
$V_{ij}$.) The dynamics of the relativistic $N$-body system is determined by
the search for the state $|\psi \rangle$ such that 
\begin{equation}
{\cal H}|\psi \rangle=\biggl \{\sum_{i=1}^{N}\frac{1}{2m_{i}}%
(p_{i}^{2}-m_{i}^{2})-\sum_{i=1}^{N}\sum_{j>i}^{N}V_{ij}-\sum_{i}\frac{W_{i} 
}{2m_{i}}\biggr \}|\psi \rangle=0.  \label{eq:Rham}
\end{equation}
Even without the many-body forces $W_{i\text{ }}$, this equation is
very difficult to solve because the potentials $V_{ij}$ depend on the
momenta $P_{ij}=p_{i}+p_{j}$ (through $x_{ij\perp })$ which are not
constants of motion for the $N-$body system; forces outside of the
$ij$ system produce time-dependent $P_{ij}.$ \ On the other hand, if
one \ uses the two-body Hamiltonians to generate basis states, then
for those states and Hamiltonians, one can regard the $P_{ij\text{ }}$
as constants of motion. \ That is, the Hamiltonian can be separated in
terms of two-body Hamiltonians plus residual interactions\ regarded as
perturbations. \ This greatly simplifies the problem.

\section{Some Applications}

The quadratic form of the momentum operators $p_{i}$ in the $N$-body
Hamiltonian Eq.\ (\ref{eq:Rham}) makes it easy to manipulate the momentum
terms to obtain the center-of-mass momentum and other relative momenta. The
potential term in the equation appears in a way similar to that in which it
appears in the non-relativistic case.

The relativistic $N$-body equation can be compared to the non-relativistic $%
N $-body equation. Introducing $\epsilon _{i}=p_{i}^{0}-m_{i}$, we have 
\begin{eqnarray}
\frac{1}{2m_{i}}(p_{i}^{2}-m_{i}^{2})=\epsilon _{i}-{\frac{{\bf p}%
_{i}^{2}-\epsilon _{i}^{2}}{2m_{i}}}. 
\end{eqnarray}
Equation (\ref{eq:Rham}\ ) becomes 
\begin{equation}
{\cal H}|\psi \rangle=\biggl \{E_{{\rm NR}}-\sum_{i=1}^{N}{\frac{{\bf p_{i}}%
^{2}-\epsilon _{i}^{2}}{2m_{i}}}-\sum_{i=1}^{N}\sum_{j>i}^{N}V_{ij}-\sum_{i}%
\frac{W_{i}}{2m_{i}}\biggr \}|\psi \rangle=0,  \label{eq:Nham}
\end{equation}
where $E_{{\rm NR}}=\sum_{i}\epsilon _{i}$. This is identical to the
non-relativistic $N$-body Hamiltonian with eigenvalue $E_{{\rm NR}}$
when $|\epsilon |_{i}<<m_{i}$ and the $W_i$ are neglected.

In the next example, we can examine a system of an even number of $N$
particles forming $N/2$ composite particles, as in a system of $N/2$ mesons.
For such a system, one can consider an initial state of the form 
\begin{equation}
|\psi _{a}\rangle=|\{(i_{1}j_{1}),(i_{2},j_{2}),...(i_{N/2}j_{N/2})\}\rangle,
\end{equation}
in which particles $i_{\alpha }$ and $j_{\alpha }$ form a composite two-body
subsystem $(i_{\alpha }j_{\alpha })$. Subsequent dynamics of the system is
determined by the evolution operator containing ${\cal H}$ and the reaction
matrix element $\langle\psi _{a}^{\prime }|{\cal H}|\psi _{a}\rangle$ where 
\begin{equation}
|\psi _{a}^{\prime }\rangle=|\{(i_{1}^{\prime }j_{1}^{\prime
}),(i_{2}^{\prime },j_{2}^{\prime }),...(i_{N/2}^{\prime }j_{N/2}^{\prime
})\}\rangle.
\end{equation}
For the evaluation of the element $\langle\psi _{a}^{\prime }|{\cal H}|\psi
_{a}\rangle$ of the Hamiltonian matrix, the $N$-body Hamiltonian can be
separated into an unperturbed Hamiltonian ${\cal H}_{0}$ and a residual
interaction $V_{I}$, 
\begin{equation}
{\cal H=H}_{0}+V_{I},
\end{equation}
where the unperturbed Hamiltonian ${\cal H}_{0}$ is 
\begin{equation}
{\cal H}_{0}={\cal H}_{i_{1}j_{1}}+{\cal H}_{i_{2}j_{2}}+..+{\cal H}%
_{i_{N/2}j_{N/2}},  \label{h1}
\end{equation}
with 
\begin{equation}
{\cal H}_{ij}=\frac{1}{2m_{i}}(p_{i}^{2}-m_{i}^{2})+\frac{1}{
2m_{j}}(p_{j}^{2}-m_{j}^{2})-V_{ij},  \label{h2}
\end{equation}
and $V_{I}$, the `prior' form of the residual interaction, is 
\begin{equation}
V_{I}=-\sum_{i=1}^{N}{\sum_{j,j>i}^{N}}^{\prime }V_{ij}-\sum_{i}\frac{W_{i}}{%
2m_{i}}  \label{h3}
\end{equation}
where the summation ${\sum_{i=1}^{N}\sum_{j,j>i}^{N}}^{\prime }$ is carried
out with the set $\{ij\}$ different from those of the composite particles $%
\{i_{1}j_{1}\}$,$\{i_{2}j_{2}\}$, $\{i_{3}j_{3}\}$, ..., and $%
\{i_{N/2}j_{N/2}\}$: 
\begin{equation}
{\sum_{i=1}^{N}\sum_{j,j>i}^{N}}^{\prime }=\sum_{i=1}^{N}\sum_{j,j>i}^{N}%
\biggr |_{\{ij\}\neq \{i_{1}j_{1}\},\{i_{2},j_{2}\},...\{i_{N/2}j_{N/2}\}}.
\label{h4}
\end{equation}
Then, since 
\begin{equation}
{\cal H}_{0}|\{(i_{1}j_{1}),(i_{2},j_{2}),...(i_{N/2}j_{N/2})\}\rangle=0,
\end{equation}
the transition matrix element $\langle\psi _{a}^{\prime }|{\cal H}|\psi
_{a}\rangle$ becomes 
\begin{equation}
\langle\psi _{a}^{\prime }|{\cal H}|\psi _{a}\rangle=\langle\psi
_{a}^{\prime }|V_{I}|\psi _{a}\rangle.
\end{equation}

The separation of ${\cal H}$ into ${\cal H}_{0}$ and $V_{I}$ is not unique,
and it is important to show that the reaction matrix element of the residual
interaction is independent of the different ways of separating out the
unperturbed Hamiltonian and the residual interaction. We can alternatively
choose the unperturbed Hamiltonian to be associated with the state $|\psi
_{a}^{\prime }\rangle=|\{(i_{1}^{\prime }j_{1}^{\prime }),(i_{2}^{\prime
},j_{2}^{\prime }),...(i_{N/2}^{\prime }j_{N/2}^{\prime })\}\rangle$ such
that 
\begin{equation}
{\cal H}_{0}^{\prime }|\{(i_{1}^{\prime }j_{1}^{\prime }),(i_{2}^{\prime
},j_{2}^{\prime }),...(i_{N/2}^{\prime }j_{N/2}^{\prime })\}\rangle=0,
\end{equation}
associating with the ``post'' form of the residual interaction $%
V_{I}^{\prime }$, 
\begin{equation}
{\cal H}={\cal H}_{0}^{\prime }+V_{I}^{\prime }.
\end{equation}
The quantities ${\cal H}_{0}^{\prime }$ and $V_{I}^{\prime }$ are defined in
a way similar to that given in Eqs. (\ref{h1}-\ref{h4}). \ The reaction
matrix element between the basis states is 
\begin{eqnarray}
\langle\psi _{a}^{\prime }|V_{I}|\psi _{a}\rangle &=&\langle\psi
_{a}^{\prime }|{\cal H}_{0}+V_{I}|\psi _{a}\rangle=\langle\psi _{a}^{\prime
}|{\cal H}|\psi _{a}\rangle=\langle\psi _{a}^{\prime }|{\cal H}_{0}^{\prime
}+V_{I}^{\prime }|\psi _{a}\rangle  \nonumber \\
&=&\langle\psi _{a}^{\prime }|V_{I}^{\prime }|\psi _{a}\rangle,
\label{theorem}
\end{eqnarray}
which indicates that the reaction matrix element is the same for the
``prior'' form or the ``post'' form of the residual interactions. It is
independent of the way in which we split up the total Hamiltonian. This
``post-prior'' equivalence guarantees the uniqueness of the reaction matrix
element and insures the usefulness of the perturbation expansion.

Using this method of separating the total Hamiltonian, the matrix element of 
${\cal H}$ between any two basis states can be evaluated. The construction
of the matrix of the total Hamiltonian will allow one to construct the
evolution operator and to follow the dynamics of the system.

\section{Scattering of two composite particles}

As another explicit example, we apply our formalism to a problem of
practical interest. Consider four particles with masses $m_{i}$ and
momentum $p_{i}$ where $i=1,2,3,4$. We have the total Hamiltonian
\begin{equation}
{\cal H}=\sum_{i=1}^{4}{\frac{1}{2m_{i}}}(p_{i}^{2}-m_{i}^{2})-
\sum_{i=1}^{4}\sum_{j,j>i}^{4}V_{ij}-\sum_{i=1}^{4}\frac{W_i}{2m_{i}}.
\label{41}
\end{equation}
Using the method of the first section, we can solve for the bound states of
mass $M_{ij}$ for the motion of particles $i$ and $j$ interacting with the
interaction $V_{ij}$, 
\begin{equation}
{\cal H}_{ij}|\psi _{ij} \rangle=\left[ {\frac{1}{2m_{i}}}(p_{i}^{2}-m_{i}^{2})+{%
\frac{1}{2m_{j}}}(p_{j}^{2}-m_{j}^{2})-V_{ij}\right] |\psi _{ij}\rangle=0.
\end{equation}

We can consider the reaction of two composite particles $A(12)$ and $B(34)$
where particles 1 and 3 are particles and particles 2 and 4 are
antiparticles as in meson-meson scattering, interacting through a pair-wise
interaction $V_{ij}$. We study the relativistic quark-interchange reaction 
\begin{equation}  \label{rea}
A(12)+B(34)\rightarrow C(14)+D(32),
\end{equation}
as a generalization of the non-relativistic case investigated by Barnes and
Swanson \cite{Bar92}. For this reaction with a momentum transfer $t$ 
\begin{equation}
t=(A-C)^{2}=m_{A}^{2}+m_{C}^{2}-2A_{0}C_{0}+2{\bf A}\cdot {\bf C},
\end{equation}
the differential cross section in the first-Born approximation is given by 
\cite{Bar92} 
\begin{equation}
{\frac{d\sigma }{dt}}={\frac{1}{64\pi s}}{\frac{\hbar ^{2}}{|{\bf p} _{A}^{c
}|^{2}}}\left| {\cal M}_{fi}\right| ^{2}
\end{equation}
where ${\cal M}_{fi}$ is 
\begin{equation}
{\cal M}_{fi}=(2\pi )^{3}\sqrt{2E_{A}^{c}2E_{B}^{c}2E_{C}^{c}2E_{D}^{c}}
~~h_{fi},
\end{equation}
and $h_{fi}$ is the reaction matrix element from the initial state $%
A(12)B(34)$ to the final state $C(14)D(32)$ initiated by the residual
interaction $V_{I}$. In the above equation, a kinematic variable with the
superscript $c$ refers to that variable evaluated in the center-of-mass
(collider) system. From the above result, $h_{fi}$ has the dimension of $1/( 
{\rm mass})^2$.

To obtain the reaction matrix element $h_{fi}$ in our relativistic
formulation, we need to split the total Hamiltonian into the nonperturbative
part ${\cal H}_{0}$ and the residual interaction $V_{I}$. This can be
carried out in two different ways. In the ``prior'' form, it is split as 
\begin{eqnarray}
{\cal H}={\cal H}_{12}+{\cal H}_{34}-V_{13}-V_{14}-V_{23}-V_{24}-%
\sum_{i=1}^{4}\frac{Wi}{2m_{i}}, 
\end{eqnarray}
where the unperturbed Hamiltonian is 
\begin{eqnarray}
{\cal H}_{0}({\rm prior})={\cal H}_{12}+{\cal H}_{34}, 
\end{eqnarray}
and the residual interaction is 
\begin{eqnarray}
V_{I}({\rm prior})=-V_{13}-V_{14}-V_{23}-V_{24}-\sum_{i=1}^{4}\frac{W_{i}}{%
2m_{i}}. 
\end{eqnarray}
The reaction matrix element is 
\begin{equation}
\label{hij}
2\pi \delta ^{4}(P_{A}+P_{B}-P_{C}-P_{D})h_{fi}({\rm prior})=-\langle \psi
_{14}\psi _{23}|V_{13}+V_{14}+V_{23}+V_{24}+\sum_{i=1}^{4}\frac{W_{i}}{2m_{i}%
}|\psi _{12}\psi _{34}\rangle .
\end{equation}
In graphic form, if we represent the interaction $V_{ij}$ by a curly line,
the first four terms in the above matrix element are represented by the four
diagrams in Fig.\ 1. The interaction takes place before the rearrangement of
the constituents.

\vspace*{4.5cm} \epsfxsize=300pt \includegraphics{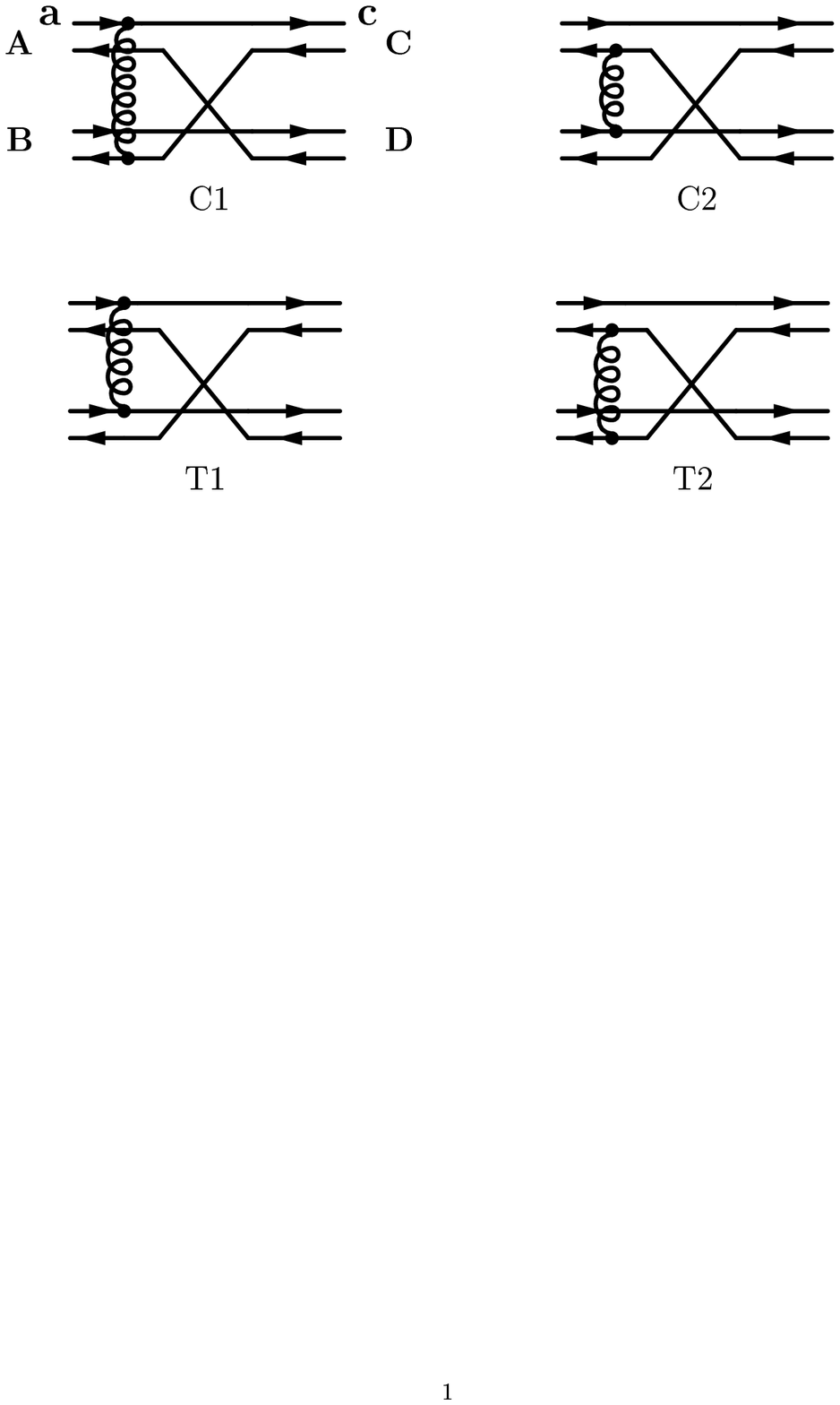} \vspace*{0.4cm}\hspace*{3cm} 
\begin{minipage}[t]{14cm}
\noindent {\bf Fig.\ 1}.  {`Prior' diagrams for the reaction A+B
$\to$ C+D.}
\end{minipage}
\vskip4truemm \noindent On the other hand, if we use the ``post'' form of
splitting the total Hamiltonian, we have 
\begin{eqnarray}
{\cal H}={\cal H}_{14}+{\cal H}_{32}-V_{13}-V_{12}-V_{43}-V_{42}-%
\sum_{i=1}^{4}\frac{W_{i}}{2m_{i}}. 
\end{eqnarray}
The unperturbed Hamiltonian is 
\begin{eqnarray}
{\cal H}_{0}({\rm post})={\cal H}_{14}+{\cal H}_{32}, 
\end{eqnarray}
and the residual interaction is 
\begin{eqnarray}
V_{I}({\rm post})=-V_{13}-V_{12}-V_{43}-V_{42}-\sum_{i=1}^{4}\frac{W_{i}}{%
2m_{i}}. 
\end{eqnarray}
The reaction matrix element is 
\begin{equation}
2\pi \delta ^{4}(P_{A}+P_{B}-P_{C}-P_{D})h_{fi}({\rm post})=-\langle \psi
_{14}\psi _{23}|V_{12}+V_{13}+V_{42}+V_{43}+\sum_{i=1}^{4}\frac{W_{i}}{2m_{i}%
}|\psi _{12}\psi _{34}\rangle .
\end{equation}
In graphic form, the first four terms in the above matrix element are
represented by the four diagrams in Fig.\ 2. The interaction takes place
after the rearrangement of the constituents.

\vspace*{4.5cm} \epsfxsize=300pt \includegraphics{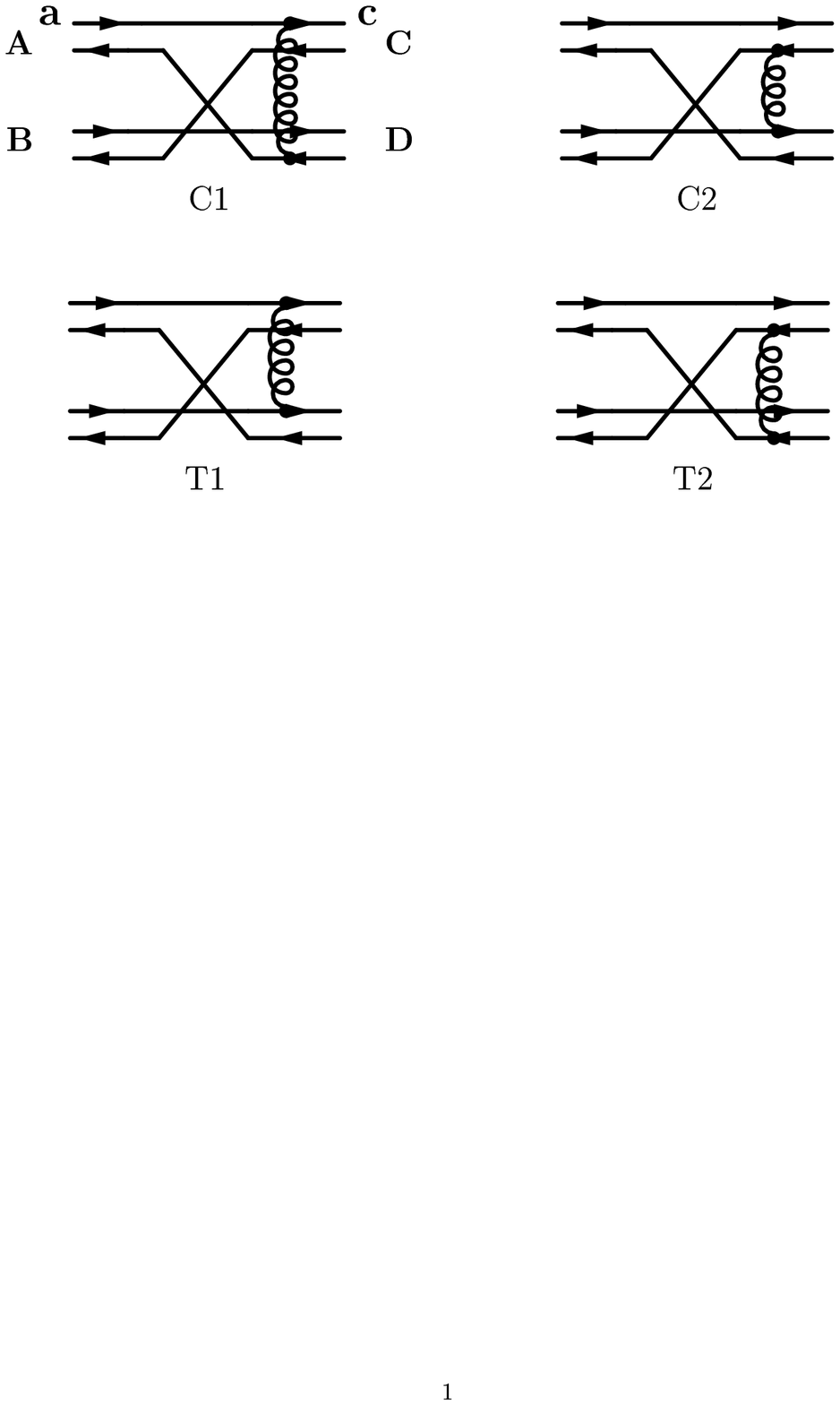} \vspace*{0.4cm}\hspace*{3cm} 
\begin{minipage}[t]{14cm}
\noindent {\bf Fig.\ 2}.  {`Post' diagrams for the reaction A+B $\to$ C+D .}
\end{minipage}
\vskip 4truemm \noindent

Therefore, if we start with the prior expression for the matrix element, we
have ${\cal H}_{12}\psi _{12}=0$ and ${\cal H}_{34}\psi _{34}=0$, and we
have (cancelling out the $N>2$-body potentials) 
\begin{eqnarray}
\langle \psi _{14}\psi _{23}|V_{13}+V_{14}+V_{23}+V_{24}|\psi _{12}\psi
_{34}\rangle &=&\langle \psi _{14}\psi _{23}|-{\cal H}_{12}-{\cal H}%
_{34}+V_{13}+V_{14}+V_{23}+V_{24}|\psi _{12}\psi _{34}\rangle  \nonumber \\
&=&\langle \psi _{14}\psi _{23}|-{\cal H}_{14}-{\cal H}%
_{23}+V_{12}+V_{13}+V_{42}+V_{43}|\psi _{12}\psi _{34}\rangle,
\end{eqnarray}
where we have used Eq.\ (\ref{41}) to write out the Hamiltonian for the
two-body system. Because ${\cal H}_{14}|\psi _{14}\rangle=0$ and ${\cal H}%
_{23}|\psi _{23}\rangle=0$, we have then 
\begin{equation}
\langle \psi _{14}\psi _{23}|V_{13}+V_{14}+V_{23}+V_{24}|\psi _{12}\psi
_{34}\rangle =\langle \psi _{14}\psi _{23}|V_{12}+V_{13}+V_{42}+V_{43}|\psi
_{12}\psi _{34}\rangle ,
\end{equation}
which leads to the relativistic generalization of the post-prior equivalence
of the reaction matrix element $h_{fi}$, 
\begin{equation}
h_{fi}({\rm prior})=h_{fi}({\rm post}).
\end{equation}
Just as in non-relativistic reaction theory \cite{Sch68}, the
equivalence is possible only when one uses the same internal relative
wave function for the composite particles in their scattering process
as in the bound state problem for the individual composite
particles. The equivalence allows a unique determination of the
reaction cross section in the first-Born approximation.

\section{Second Quantization of Particles in a Bound State}

A practical problem arises when one attempts to evaluate the reaction matrix
element $\langle \psi _{a}|V_{I}|\psi _{a}^{\prime }\rangle $, 
\begin{eqnarray}
\langle \psi _{a}|V_{I}|\psi _{a}^{\prime }\rangle =\langle
(i_{1}j_{1}),(i_{2}j_{2}),...|-\sum_{i=1}^{N}{\sum_{j,j>i}^{N}}^{\prime
}V_{ij}-\sum_{i}{\frac{W_{i}}{2m_{i}}}|(i_{1}^{\prime }j_{1}^{\prime
}),(i_{2}^{\prime }j_{2}^{\prime }),...\rangle,
\end{eqnarray}
where $\{(i_{1}j_{1}),(i_2 j_2),...\}$ represents composite two-body
subsystems. For a pair-wise interaction $V_{ij}$ in the above
equation, we have
\begin{equation}
\langle \psi _{a}|V_{ij}|\psi _{a}^{\prime }\rangle =\langle (i\alpha
^{\prime }),(j\beta ^{\prime })|V_{ij}|(i\alpha ),(j\beta )\rangle ,
\end{equation}
where $\{\alpha ^{\prime }\beta ^{\prime }\}$ is a permutation of $\{\alpha
\beta \}$. The composite wave functions $\psi _{i\alpha }$ are usually
computed in the CM of the $(i\alpha )$ composite particle system (the $q_{i} 
\bar{q}_{\alpha }$ system in our example of a system of mesons). However,
this is not the same as the so-called collider frame (the CM frame of the $%
(i\alpha )$-$(j\beta )$ meson-meson system) either in the bra or ket states.
For non-relativistic reactions, the relative wave function of a composite
system in the collider frame is obtained from the wave function for the
composite particle at rest by a Galilean boost, and they are related by a
simple momentum shift. In the relativistic case, a Lorentz boost is needed
in place of a Galilean boost. \ We need, therefore, to discuss the Lorentz
transformation of the state of the composite system.

We represent the state of the composite two-body system with a 4-momentum $P$
by
\begin{eqnarray}
\langle P^{0}|(12)_{P}\rangle &=&4\varepsilon _{1}\varepsilon _{2}\int
d^{4}p_{1}d^{4}p_{2}\delta ^{4}(p_{1}+p_{2}-P)\theta (p_{1}^{0})\theta
(p_{2}^{0})  \nonumber \\
&&\times \delta (p_{1}^{2}-m_{1}^{2}-\Phi (x_{\perp }))\delta
(p_{2}^{2}-m_{2}^{2}-\Phi (x_{\perp })){\psi }(p_{1},p_{2})|p_{1}p_{2}
\rangle ,  \label{20}
\end{eqnarray}
where we use the same symbol $p_{1}$ and $p_{2}$ to denote $c$-numbers
and operators, using the context to distinguish between them.  The
factor of $4\varepsilon _{2}\varepsilon _{2}$ is included so that we
obtain the usual results in the nonrelativistic limit.  The delta
function containing the composite particle energy $P^0$ arises from the
projection of the energy eigenstate state $|P^0\rangle$ onto the state
vector $|(12)_{P}\rangle $ where $P=\{E,{\bf P}\}$ and $E=\sqrt{{\bf
P}^2+M^2}$ (see also Eq.\ (\ref{p0}) below).  Hence,
we use the notation $ \langle P^{0}|(12)_{{E}{\bf P}}\rangle $ to
denote the projection of the state vector $|(12)_{P}\rangle $ onto the
energy eigenstate state $|P_0\rangle$.  The above state is constructed
in analogy to the two free-particle states
\begin{equation}
\langle P^{0}|(12)_{P}\rangle =4\varepsilon _{1}\varepsilon _{2}\int
d^{4}p_{1}d^{4}p_{2}\delta ^{4}(p_{1}+p_{2}-P)\theta (p_{1}^{0})\theta
(p_{2}^{0})\delta (p_{1}^{2}-m_{1}^{2})\delta (p_{2}^{2}-m_{2}^{2}){\psi }%
(p_{1},p_{2})|p_{1}p_{2}\rangle .  \label{eq:P2}
\end{equation}
However, unlike the free-particle state, the state (\ref{20})
satisfies the simultaneous constraint conditions of Eq. (\ref{1})
\begin{equation}
{\cal H}_{i}|(12)_{P}\rangle =[p_{i}^{2}-m_{i}^{2}-\Phi (x_{\perp
})]|(12)_{P}\rangle =0{~~~~~{\rm for~~~~~}}i=1,2,
\end{equation}
so that neither particle is on mass shell. The definition (\ref{20})
also introduces a momentum space wave function ${\psi }(p_{1},p_{2})$
defined so that it has positive constituent energies. We emphasize
that the momentum eigenstates $|p_{1}p_{2}\rangle $ in this expansion
are off shell.  That is
\begin{equation}
(p_{i}^{2}-m_{i}^{2})|p_{1}p_{2}\rangle \neq 0.
\end{equation}

Our first step is to show that the above bound state composite is a sharp
state being zero unless $P^{2}=M^{2}$ where $M$ is the meson bound state
mass. By using the total and relative momentum operators and Eq. (\ref{eps})
(so that Eq. (\ref{eqpq}) is satisfied and not $P\cdot q\equiv 0$), the
product of the two delta functions can be written as 
\begin{eqnarray}
&&\delta (p_{1}^{2}-m_{1}^{2}-\Phi (x_{\perp }))\delta
(p_{2}^{2}-m_{2}^{2}-\Phi (x_{\perp }))  \nonumber \\
&=&\delta (\frac{\varepsilon _{1}^{2}P^{2}}{M^{2}}+q^{2}-m_{1}^{2}-\Phi
(x_{\perp })+\frac{2\varepsilon _{1}P\cdot q}{M})\delta
((m_{1}^{2}-m_{2}^{2})\frac{(P^{2}-M^{2})}{M^{2}}+2P\cdot q)  \nonumber \\
&=&\delta (\frac{\varepsilon _{1}\varepsilon _{2}(P^{2}-M^{2})}{M^{2}}+{\cal %
H}_{q}{\cal )}\delta ((m_{1}^{2}-m_{2}^{2})\frac{(P^{2}-M^{2})}{M^{2}}%
+2P\cdot q)
\end{eqnarray}
where ${\cal H}_{q}=q^{2}-\Phi (x_{\perp })+b^{2}$. \ We assume that the
momentum space wave function ${\psi }(p_{1},p_{2})$ is an
eigenfunction of ${\cal H}_{q}$, so that ${\cal H}_{q}{\ \psi }%
(p_{1},p_{2})=0 $. Thus, on such states the above equation becomes 
\begin{equation}
\delta (p_{1}^{2}-m_{1}^{2}-\Phi (x_{\perp }))\delta
(p_{2}^{2}-m_{2}^{2}-\Phi (x_{\perp }))=\frac{M^{2}}{2\varepsilon
_{1}\varepsilon _{2}}\delta (P^{2}-M^{2}{\cal )}\delta (P\cdot q).
\end{equation}
This shows, as anticipated, that the state defined in Eq. (\ref{20})
is sharp and would satisfy Eq. (\ref{pm}) in addition to Eq. (\ref{1}).

We allow the delta function arguments to operate on the momentum states, and
use the positive energy condition. Then from Eq. (\ref{qrel}) we have 
\begin{eqnarray}  \label{6p9}
&&\theta (p_{1}^{0})\theta (p_{2}^{0})\delta (p_{1}^{2}-m_{1}^{2}-\Phi
(x_{\perp }))\delta (p_{2}^{2}-m_{2}^{2}-\Phi (x_{\perp }))  \nonumber \\
&=&\theta (p_{1}^{0})\theta (p_{2}^{0})\frac{M^{2}}{4\varepsilon
_{1}\varepsilon _{2}E^{2}}\delta (p_{1}^{0}-\frac{E}{M}\varepsilon _{1}-{\bf %
P\cdot q)}\delta (p_{2}^{0}-\frac{E}{M}\varepsilon _{2}+{\bf P\cdot q)},
\end{eqnarray}
where $E=\sqrt{{\bf P}^{2}+M^{2}}.$ \ We would like to express this in terms
of a Lorentz transformation from the CM system (in which the meson has a
mass $M$)$.$ \ 

When the composite particle is boosted by $\Lambda $ to a momentum ${\bf P}$%
, the meson's velocity is 
\begin{equation}
{\bf V}=\frac{{\bf P}}{E}.
\end{equation}
Thus with $\hat{u}={\bf \hat{V}=P}/|{\bf P}|$, the components of $\Lambda $
are 
\begin{eqnarray}
\Lambda _{k}^{i} &=&\delta _{ik}+(\gamma -1)\hat{u}_{i}\hat{u}_{k}=\delta
_{ik}+(\frac{E}{M}-1)\frac{{\bf P}_{i}{\bf P}_{k}}{{\bf P}^{2}},  \nonumber
\\
\Lambda _{0}^{i} &=&\Lambda _{i}^{0}=\hat{u}_{i}\sqrt{(\gamma ^{2}-1)}=\frac{%
{\bf P}_{i}}{M},  \nonumber \\
\Lambda _{0}^{0} &=&\gamma =\frac{E}{M}.  \label{lt}
\end{eqnarray}
We use the notation $p_{1}^{\ast }=(\varepsilon _{1},{\bf p}_{1}^{\ast })$
to represent the four-momenta of the $i$th constituent in the composite
particle rest frame, and $p_{i}=p_{i\Lambda }=(\varepsilon _{i\Lambda },{\bf %
p}_{i\Lambda })$ to represent the 4-momentum of the $i$th constituent in the
frame boosted by $\Lambda $. Then we have 
\begin{equation}
p_{i}^{0}\equiv p_{i\Lambda }^{0}=(\Lambda p_{1}^{\ast })^{0}=\gamma
p_{i}^{\ast 0}+\hat{u}\cdot {\bf p}_{i}^{\ast }=\frac{E\varepsilon _{i}}{M}+%
\frac{{\bf P}\cdot {\bf p}_{i}^{\ast }}{M}\equiv \varepsilon _{i\Lambda
},\,\,\,\ \ \ \,\,(i=1,2)
\end{equation}
and 
\begin{equation}
{\bf p}_{i}\equiv {\bf p}_{i\Lambda }=\frac{{\bf P}}{M}\varepsilon _{i}+{\bf %
p}_{i}^{\ast }+(\frac{E}{M}-1)\frac{{\bf P}{\bf P}\cdot {\bf p}_{i}^{\ast }}{%
{\bf P}^{2}}\,,\ \ \ \ \ \ \ \ \ \,\,(i=1,2).  \label{space}
\end{equation}
Because ${\bf p}_{1}^{\ast }+{\bf p}_{2}^{\ast }=0$ we have therefore 
\begin{equation}
p_{1}^{0}+p_{2}^{0}=\sqrt{{\bf P}^{2}+M^{2}}=E,
\end{equation}
\begin{equation}
{\bf p}_{1}+{\bf p}_{2}=\frac{{\bf P}}{M}(\varepsilon _{1}+\varepsilon _{2})=%
{\bf P}.
\end{equation}
Furthermore, using Eq.(\ref{qrel}) we have 
\begin{equation}
{\bf P\cdot q=}\frac{E}{M}{\bf P\cdot q}^{\ast }=\frac{E}{M}{\bf P\cdot p}%
_{1}^{\ast }=-\frac{E}{M}{\bf P\cdot p}_{2}^{\ast }.
\end{equation}
Hence, Eq.\ (\ref{6p9}) can be rewritten as 
\begin{eqnarray}
&&\theta (p_{1}^{0})\theta (p_{2}^{0})\delta (p_{1}^{2}-m_{1}^{2}-\Phi
(x_{\perp }))\delta (p_{2}^{2}-m_{2}^{2}-\Phi (x_{\perp }))  \nonumber \\
&=&\theta (p_{1}^{0})\theta (p_{2}^{0})\frac{M^{2}}{4\varepsilon
_{1}\varepsilon _{2}E^{2}}\delta (p_{1}^{0}-(\Lambda p_{1}^{\ast
})^{0})\delta (p_{2}^{0}-(\Lambda p_{2}^{\ast })^{0}).
\end{eqnarray}
We have therefore 
\begin{equation}
\langle P^{0}|(12)_{E{\bf P}}\rangle ={\frac{M^{2}}{E^{2}}}\int
d^{4}p_{1}d^{4}p_{2}\delta ^{4}(P-p_{1}-p_{2})\delta (p_{1}^{0}-(\Lambda
p_{1}^{\ast })^{0})\delta (p_{2}^{0}-(\Lambda p_{2}^{\ast })^{0}){\psi }%
(p_{1},p_{2})|p_{1},p_{2}\rangle .
\end{equation}
Now the $p_{1}^{0}$ and $p_{2}^{0}$ part of the $d^{4}p_{1}$ and $d^{4}p_{2}$
can be integrated out and the result is 
\begin{equation}
\langle P^{0}|(12)_{E{\bf P}}\rangle =\frac{M^{2}}{E^{2}}\delta
(P^{0}-p_{1}^{0}-p_{2}^{0})\int d{\bf p}_{1}d{\bf p}_{2}\delta ^{3}({\bf P}-%
{\bf p}_{1}-{\bf p}_{2}){\psi }(p_{1}^{0}{\bf p}_{1},p_{2}^{0}{\bf p}%
_{2})|p_{1}^{0}{\bf p}_{1},p_{2}^{0}{\bf p}_{2}\rangle ,  \label{dlt1}
\end{equation}
where $p_{i}^{0}=(\Lambda p_{i}^{\ast })^{0}$ and $p_1^0+p_2^0=E$. In
the CM system, $ p_{1}^{0}+p_{2}^{0}=M$ and the state vector is
\begin{equation}
\langle P^{0}|(12)_{M{\bf 0}}\rangle =\delta (P_{0}-M)\int d{\bf p}%
_{1}^{\ast }d{\bf p}_{2}^{\ast }\delta ^{3}({\bf p}_{1}^{\ast }+{\bf p}%
_{2}^{\ast }){\psi }_{M}(\varepsilon _{1}{\bf p}_{1}^{\ast },\varepsilon _{2}%
{\bf p}_{2}^{\ast })|\varepsilon _{1}{\bf p}_{1}^{\ast },\varepsilon _{2}%
{\bf p}_{2}^{\ast }\rangle .  \label{619}
\end{equation}
We introduce the notation $|M({\bf P})\rangle $ defined as 
\begin{equation}
|M({\bf P})\rangle =\int d{\bf p}_{1}d{\bf p}_{2}\delta ^{3}({\bf P}-{\bf p}%
_{1}-{\bf p}_{2}){\psi }(p_{1}^{0}{\bf p}_{1},p_{2}^{0}{\bf p}_{2})|p_{1}^{0}%
{\bf p}_{1},p_{2}^{0}{\bf p}_{2}\rangle ,
\end{equation}
so that, since in a general frame $p_{1}^{0}+p_{2}^{0}=\sqrt{{\bf
P}^{2}+M^{2}}=E,$
\begin{equation}
\langle P^{0}|(12)_{E{\bf P}}\rangle =\frac{M^{2}}{E^{2}}\delta
(P^{0}-\sqrt{{\bf P}^{2}+M^{2}})|M({\bf P})\rangle .  \label{p0}
\end{equation}
The projection of the state vector $|(12)_{E{\bf P}}\rangle $ onto the
time component of the center of mass \cite{time} is then 
\begin{eqnarray}
\langle t|(12)_{E{\bf P}}\rangle =\int \langle t|P^{0}\rangle
dP^{0}\langle P^{0}|(12)_{E{\bf P}}\rangle   
=e^{-iEt}\frac{M^{2}}{E^{2}}|M({\bf P}
)\rangle .  \label{time}
\end{eqnarray}

From the above results, the energies of the constituents in a composite
particle take on fixed values ($p_{i}^{0\ast }=\varepsilon _{1}$) in the
center-of-mass system, while their off-shell component ${\bf p}_{i}^{\ast }$
takes on continuous variations with a distribution. On the other hand, when
boosted by the Lorentz transformation $\Lambda $, the energy of the $i$th
constituent is given by $p_{i}^{0}=(\Lambda p_{i}^{\ast })^{0}$ in a moving
composite particle, but their sum, $p_1^0+p_2^0$, remains a constant. Even
though the time-like components of the constituents have these well-defined
values which depend on the frame of reference, they are often not written
out explicitly, for brevity of notation.

For a proper treatment of spin, we should parallel the treatment in the
above sections except using Dirac operators instead of Klein-Gordon
operators. Alternatively, we can adapt the above spinless results to the
case of spin by using the fact that we can reduce, for two particles, the
two-body Dirac equations to Schr\"{o}dinger-like forms above but with $\Phi $
depending on spin degrees of freedom\cite{cra88},\cite{cra94}. In a future
paper we shall include the spin dependent features in more detail.

In order to deal with multiparticle configurations, we introduce creation
and annihilation operators of the constituents in a composite particle in
its CM frame 
\begin{equation}
|{\bf p}_{1}^{\ast },{\bf p}_{2}^{\ast }\rangle=b^{\dagger }({\bf p}%
_{1}^{\ast })d^{\dagger }({\bf p}_{2}^{\ast })|0\rangle.
\end{equation}
(We suppress spin, flavor, and color indices.) We assume the general
expression 
\begin{eqnarray}
\{b^{\dagger }({\bf p}_{1}^{\prime \ast }),b({\bf p}_{1}^{\ast })\} &=&N(%
{\bf p}_{1}^{\ast })\delta ({\bf p}_{1}^{\prime \ast }-{\bf p}_{1}^{\ast }) 
\nonumber \\
\{d^{\dagger }({\bf p}_{2}^{\prime \ast }),d({\bf p}_{2}^{\ast })\} &=&N(%
{\bf p}_{2}^{\ast })\delta ({\bf p}_{2}^{\prime \ast }-{\bf p}_{2}^{\ast })
\end{eqnarray}
and thus 
\begin{equation}
\langle{\bf p}_{1}^{\prime \ast },{\bf p}_{2}^{\prime ^{\ast }}|{\bf p}%
_{1}^{\ast },{\bf p}_{2}^{\ast }\rangle=N({\bf p}_{1}^{\ast })\delta ({\bf p}%
_{1}^{\prime \ast }-{\bf p}_{1}^{\ast })N({\bf p}_{2}^{\ast })\delta ({\bf p}%
_{2}^{\prime \ast }-{\bf p}_{2}^{\ast }).
\end{equation}
In the case in which free isolated particles are created and
annihilated, one traditionally takes either $N({\bf p}^{\ast })=1$ or
$N({\bf p}^{\ast })=2\sqrt{{\bf p}^{\ast 2}+m^{2}}$. We emphasize,
however, that the above momentum in the creation and annihilation
operators are not on mass shell but on energy shell. In the context of
the constraint approach, the individual creation and annihilation
operators do not produce free-particle states, but rather constituent
states within a composite associated with a definite total mass and
total momentum. Since the aim of this paper is a description of the
relativistic $N-$body problem in a separable two-body basis, this is
plausible. To achieve this, we must determine how the creation and
annihilation operators will transform under a Lorentz
transformation. Let $U(\Lambda )$ be our unitary boost operator
defined so that
\begin{equation}
U(\Lambda )b^{\dagger }({\bf p}_{1}^{\ast })U^{-1}(\Lambda )=C({\bf p}
_{1\Lambda })b^{\dagger }({\bf p}_{1\Lambda })
\end{equation}
and 
\begin{equation}
U(\Lambda )d^{\dagger }({\bf p}_{2}^{\ast })U^{-1}(\Lambda )=C({\bf p}
_{2\Lambda })d^{\dagger }({\bf p}_{2\Lambda })
\end{equation}
where{\bf \ }${\bf p}_{i\Lambda }$ is the three-vector part of $\Lambda
p_{i}^{\ast }$. In\ the case in which free isolated particles are produced,
the above two conventions lead respectively to either $C({\bf p})=\sqrt{%
(\Lambda p)^{0}/p^{0}}$ or $\ C({\bf p})=1$. In order to see what these
factors become now in the constraint approach, we consider 
\begin{eqnarray}
U(\Lambda )\{b^{\dagger }({\bf p}_{1}^{\prime \ast }),b({\bf p}_{1}^{\ast
})\}U^{-1}(\Lambda ) &=&N({\bf p}_{1}^{\ast })\delta ({\bf p}_{1}^{\prime
\ast }-{\bf p}_{1}^{\ast })  \nonumber \\
&=&C({\bf p}_{1\Lambda })C^{\ast }({\bf p}_{1\Lambda }^{\prime })N({\bf p}%
_{1\Lambda })\delta ({\bf p}_{1\Lambda }^{\prime }-{\bf p}_{1\Lambda }).
\label{dlta}
\end{eqnarray}
To make use of this equation, we need to express the boosted momenta
in terms of \ the unboosted momenta. The Lorentz transformation matrix
$\Lambda $, defined in Eq.\ (\ \ref{lt}), is independent of the
momentum of the system being boosted. For the same two-body system,
the space-like part (Eq.\ (\ref {space}) of the Lorentz transformation
on two different momenta (for the same quark, that is, $\varepsilon
_{1}=\varepsilon _{1}^{\prime }$) yields
\begin{equation}
{\bf p}_{1\Lambda }-{\bf p}_{1\Lambda }^{\prime }=[{\bf 1}+(\frac{E}{M}-1) 
\frac{{\bf P}{\bf P}}{{\bf P}^{2}}]({\bf p}_{1}^{\ast }-{\bf p}_{1}^{\ast
\prime }).  \label{popp}
\end{equation}
Using the fact that $\delta ^{3}(A{\bf r})=\delta ^{3}({\bf r})/\det |A|$,
we find that 
\begin{equation}
\delta ({\bf p}_{1\Lambda }-{\bf p}_{1\Lambda }^{\prime })=\frac{M}{E}\delta
({\bf p}_{1}^{\ast \prime }-{\bf p}_{1}^{\ast })  \label{pdpp}
\end{equation}
so that Eq.(\ref{dlta}) becomes 
\begin{equation}
N({\bf p}_{1}^{\ast })=C({\bf p}_{1\Lambda })C^{\ast }({\bf p}_{1\Lambda
}^{\prime })N({\bf p}_{1\Lambda })\frac{M}{E}.
\end{equation}
We choose the normalization $N=1$. Using phase convention with real $C$, the
simplest choice is 
\begin{equation}
C=\sqrt{\frac{E}{M}}.
\end{equation}
This $C$ is associated with the motion of the composite particle. Thus we
have\ 
\begin{equation}
U(\Lambda )b^{\dagger }({\bf p}_{1}^{\ast })U^{-1}(\Lambda )=\sqrt{\frac{E}{M%
}}b^{\dagger }({\bf p}_{1\Lambda })  \label{bbst}
\end{equation}
and 
\begin{equation}
U(\Lambda )d^{\dagger }({\bf p}_{2}^{\ast })U^{-1}(\Lambda )=\sqrt{\frac{E}{M%
}}d^{\dagger }({\bf p}_{2\Lambda }).  \label{dbst}
\end{equation}
(As anticipated above, this contrasts with the on mass shell factor given in
standard texts (see \cite{wnbrg} Eq.(4.2.12)) which refer only to the
constituent momenta.) Thus we have 
\begin{eqnarray}
|M({\bf P}) &\rangle&=U(\Lambda )|M({\bf 0})\rangle  \nonumber \\
&=&\int d^{3}p_{1}^{\ast }d^{3}p_{2}^{\ast }\psi _{M}({\bf p}_{1}^{\ast }, 
{\bf p}_{2}^{\ast })\delta ({\bf p}_{1}^{\ast }+{\bf p}_{2}^{\ast })\frac{E}{
M}b^{\dagger }({\bf p}_{1\Lambda })d^{\dagger }({\bf p}_{2\Lambda
})|0\rangle.
\end{eqnarray}
We emphasize that the transformation equations (Eq.\ (\ref{bbst}) and
Eq.\ (\ref {dbst})) are valid for arbitrary ${\bf p}_{1}$ or ${\bf p}_{2}$,
not just ones that satisfy the rest condition of ${\bf p}_{1}+{\bf
p}_{2}=0$. This implies that the creation and annihilation operators
that go into making up the individual two-body interactions will
transform in a similar way. (See Eq.(\ref{mtrx}) below.)

Next we change variables so that the delta function reflects the new total
momentum ${\bf P}$. \ The inverse of the above Lorentz transformation gives
for $i=1,2$ 
\begin{eqnarray}
{\bf p}_{i}^{\ast } &=&-\frac{{\bf P}\varepsilon _{i}}{E}+{\bf p}_{i\Lambda
}+(\frac{M}{E}-1)\frac{{\bf P}{\bf P}\cdot {\bf p}_{i\Lambda }}{{\bf P}^{2}}
\label{bst}
\end{eqnarray}
which in turn gives 
\begin{eqnarray}
(\Lambda p_{i}^{\ast })^{0} &=&\frac{\varepsilon _{i}M}{E}+\frac{{\bf P}
\cdot {\bf p}_{i\Lambda }}{E}.
\end{eqnarray}
Note that with ${\bf p}_{1\Lambda }+{\bf p}_{2\Lambda }={\bf P}$ we have 
\begin{equation}
{\bf p}_{1}^{\ast }+{\bf p}_{2}^{\ast }={\bf p}_{1\Lambda }+{\bf p}%
_{2\Lambda }-{\bf P}=0.
\end{equation}
Computing the Jacobian of the above transformation (\ref{bst}) gives 
\begin{equation}
d^{3}p_{i}^{\ast }=\frac{M}{E}d^{3}p_{i\Lambda }. 
\end{equation}
(Note again how this contrasts with the case of free on-shell
particles where $d^{3}p_{i}=d^{3}p_{i\Lambda }{p_{i}^{0}/}{(\Lambda
p_{i})^{0}}.)$ Hence, \ 
\begin{equation}
|M({\bf P)}\rangle=\frac{M}{E}\int d^{3}p_{1\Lambda }d^{3}p_{2\Lambda }\psi
_{M}( {\bf p}_{1}^{\ast },{\bf p}_{2}^{\ast })\delta ({\bf P}-{\bf p}%
_{1\Lambda }- {\bf p}_{2\Lambda })b^{\dagger }({\bf p}_{1\Lambda
})d^{\dagger }({\bf p} _{2\Lambda })|0\rangle.
\end{equation}
The wave function $\psi _{M}({\bf p}_{1}^{\ast },{\bf p}_{2}^{\ast })$
is actually $\psi _{M}({\bf p}_{1}^{\ast }({\bf p}_{1\Lambda }),{\bf
p} _{2}^{\ast }({\bf p}_{2\Lambda }))$ where the vectors $\{{\bf
p}_{1}^{\ast }({\bf p} _{1\Lambda }),{\bf p}_{2}^{\ast }({\bf
p}_{2\Lambda })\}$ are  $\{({\bf p}_{1\Lambda })_{\Lambda
^{-1}},({\bf p}_{2\Lambda })_{\Lambda ^{-1}}\}$ .\ A simple relabeling
(not a transformation) gives the representation of a general state for
a composite particle with momentum ${\bf P}$,
\begin{eqnarray}
\label{MP}
|M({\bf P})\rangle=\frac{M}{E}\int d^{3}p_{1}d^{3}p_{2}\psi _{M}({\bf p}
_{1}{}_{\Lambda ^{-1}},{\bf p}_{2}{}_{\Lambda ^{-1}}) 
 \delta ({\bf P}-{\bf p} _{1}-{\bf p}_{2})b^{\dagger }({\bf p}%
_{1})d^{\dagger }({\bf p}_{2})|0\rangle.
\end{eqnarray}
In the nonrelativistic limit, this becomes 
\begin{eqnarray}
|M({\bf P})\rangle&=&\int d^{3}p_{1}d^{3}p_{2}\psi _{M}({\bf p}_{1}{}-\frac{
m_{1}}{ (m_{1}+m_{2})}{\bf P,p}_{2}{}-\frac{m_{2}}{(m_{1}+m_{2})}{\bf P}) 
\nonumber \\
&\times& \delta ( {\bf P}-{\bf p}_{1}-{\bf p}_{2})b^{\dagger }({\bf p}
_{1})d^{\dagger }({\bf p} _{2})|0\rangle.
\end{eqnarray}
So, each composite state, Eq.\ (\ref{p0}) with Eq.\ (\ref{MP}),
differs from that of the nonrelativistic limit by not only replacing
Galilean boosts with inverse Lorentz boosts but also by a factor of the ratio
of the total CM energy to the lab energy.

In Appendix A we show that the scalar product for two systems in the same
internal state is given by the three-dimensional momentum delta function
times a covariant form 
\begin{equation}
\langle M({\bf P}^{\prime })|M({\bf P})\rangle =\frac{\delta ^{3}({\bf P}%
^{\prime }-{\bf P})M^{3}}{E^{2}}\int d^{4}p\delta (p\cdot P)|\psi
_{M}(p)|^{2},
\end{equation}
with the wave function having the same dimensions as in the nonrelativistic
case.

Given these preliminaries, we consider how to use this formulation in the
calculation of meson-meson scattering amplitudes. In this problem one starts
with a state $|(12)(34)\rangle $ consisting of two quark-antiquark states. \
We model the interaction by the exchange of an (effective) gluon
corresponding to $V$. \ At lowest order, the exchange could not produce a
final state $|(12)(34)\rangle $ but only $|(14)(23)\rangle $ since the
emission of a virtual gluon would leave the resultant initial state as two
color octet mesons rather than singlet mesons. \ One would thus need to
evaluate a typical matrix element of the form 
\begin{equation}
\langle (14)(23)|V(\hat{x}_{14\perp })|(12)(34)\rangle .
\end{equation}
Inserting $\int dt|t\rangle \langle t|=1$ and using Eq. (\ref{time})
into the above expression, we can carry out the integration in $t$ to
obtain a delta function which describes the condition of total energy
conservation. We have
\begin{eqnarray}
\label{v14}
\langle (14)(23)|V(\hat{x}_{14\perp })|(12)(34)\rangle  &=&2\pi \delta
(E_{12}+E_{34}-E_{13}-E_{24})\left( \frac{M_{12}M_{14}M_{13}M_{24}}{%
E_{12}E_{14}E_{13}E_{24}}\right) ^{2}  \nonumber \\
&&\times \langle M({\bf P}_{14}),M({\bf P}_{23})|V(\hat{x}_{14\perp })|M(
{\bf P}_{12}),M({\bf P}_{34})\rangle 
\end{eqnarray}
where
\begin{equation}
|M({\bf P}_{12})\rangle =\frac{M_{12}}{E_{12}}\int d^{3}p_{1}d^{3}p_{2}\psi (
{\bf p}_{1}{}_{\Lambda _{12}^{-1}},{\bf p}_{2}{}_{\Lambda _{12}^{-1}})\delta
({\bf P}_{12}-{\bf p}_{1}-{\bf p}_{2})b^{\dagger }({\bf p}_{1})d^{\dagger }(
{\bf p}_{2})|0\rangle ,
\end{equation}
in which ${\bf P}_{12}$ is the momentum of the composite with CM energy $
M_{12}$ so that $E_{12}=\sqrt{{\bf P}_{12}^{2}+M_{12}^{2}}$ and $\Lambda
_{12}^{-1}$ is the inverse boost to the rest system of the composite.\ Similar
expressions appear for $|M({\bf P}_{34})\rangle $, $|M({\bf P}_{14})\rangle $
, and $|M({\bf P}_{23})\rangle $.

\section{Reaction Matrix Element}

In order to compute matrix elements of the potential, we need its
second quantized version. \ We evaluate it in the rest frame of the
two interacting constituents and we assume that the second quantized
form of the potential has the same relation to its first quantized
form as in the nonrelativistic case. For the reaction
$A(12)+B(34)\rightarrow C(14)+D(32)$, the matrix elements for the
interaction between a particle and an antiparticle consists of the C1
and the C2 diagram in Fig.\ 1. We consider the C1 diagram as a
representative case.  The interaction corresponding to this C1 diagram
takes place between particles 1 and 4 and is given by
\begin{eqnarray}
& &V({\hat{x}}_{14\perp })|_{{\bf P}_{14}=0}=V({\bf \hat{x}}_{14})=\int
d^{3}x_{1}^{\prime }d^{3}x_{4}^{\prime }d^{3}x_{1}^{\prime \prime
}d^{3}x_{4}^{\prime \prime }V({\bf x}_{14}^{\prime })\delta ({\bf x}
_{1}^{\prime }-{\bf x}_{1}^{\prime \prime })\delta ({\bf x}_{4}^{\prime }-
{\bf x}_{4}^{\prime \prime })|{\bf x}_{1}^{\prime }{\bf x}_{4}^{\prime
}\rangle \langle {\bf x}_{1}^{\prime \prime }{\bf x}_{4}^{\prime \prime }| 
\nonumber \\
&=&\int d^{3}p_{1}d^{3}p_{4}d^{3}p_{1}^{\prime }d^{3}p_{4}^{\prime }\delta (
{\bf p}_{1}+{\bf p}_{4}-{\bf p}_{1}^{\prime }-{\bf p}_{4}^{\prime })\tilde{V}
({\bf p}_{1}-{\bf p}_{4}^{\prime })b^{\dagger }({\bf p}_{1})d^{\dagger }(
{\bf p}_{4})d({\bf p}_{4}^{\prime })b({\bf p}_{1}^{\prime }),
\end{eqnarray}
in which the integrals include sums over spin, flavor, and color. In this
form the indices and primes serve to lable the quark color and flavor as
well as the momentum. The annihilation and creation operators with momentum $%
p_{i}$ and $p_{i}^{\prime }$ apply only to particle $i$. To represent the
interaction $V({\bf \hat{x}}_{14})$, we choose to represent the annihialtion
and creation operators in the rest frame of the (paticle 1)-(particle 4)
pair, i.e. the (${\bf P}_{14}=0$) frame. In this frame, let the total
momenta of the two initial meson composite systems be ${\bf P}=$ ${\bf P}%
_{12}$ $+{\bf P}_{34}$, with ${\bf P}_{12}$ and ${\bf P}_{34}$ to be the
individual incoming momenta of the two meson composite systems. The total
energy of the two meson system is given in terms of the respective CM
energies of the composite particles 
\begin{equation}
\sqrt{s}=E_{12}+E_{34}=\sqrt{{\bf P}_{12}^{2}+M_{12}^{2}}+\sqrt{{\bf P}%
_{34}^{2}+M_{34}^{2}}.
\end{equation}
The total momentum ${\bf P}$ of the two (composite) particle system is
conserved in the scattering process so that we can label the matrix element
as 
\begin{equation}
\langle M({\bf P}_{14}),M({\bf P}_{23});{\bf P}|V({\bf \hat{x}}_{14})|M({\bf %
P}_{12}),M({\bf P}_{34});{\bf P}\rangle .
\end{equation}
Let us evaluate the matrix element in the ``collider frame'' defined by 
\begin{equation}
{\bf P}_{12}^{c}+\,\,\,{\bf P}_{34}^{c}=0.
\end{equation}
The Lorentz boost to that frame is given by 
\begin{eqnarray}
\Lambda _{k}^{i} &=&\delta _{ik}+(\frac{\sqrt{{\bf P}^{2}+s}}{\sqrt{s}}-1)%
\frac{{\bf P}_{i}{\bf P}_{k}}{{\bf P}^{2}}  \nonumber \\
\Lambda _{0}^{i} &=&\Lambda _{i}^{0}=-\frac{{\bf P}_{i}}{\sqrt{s}}  \nonumber
\\
\Lambda _{0}^{0} &=&\gamma =\frac{\sqrt{{\bf P}^{2}+s}}{\sqrt{s}}
\end{eqnarray}
and takes us to 
\begin{eqnarray}
&&\langle M({\bf P}_{14}),M({\bf P}_{23});{\bf P}|V({\bf \hat{x}}_{14})|M(%
{\bf P}_{12}),M({\bf P}_{34});{\bf P}\rangle   \nonumber \\
&=&\langle M({\bf P}_{14}^{c}),M({\bf P}_{23}^{c});{\bf 0}|U(\Lambda )V({\bf 
\hat{x}}_{14})U^{-1}(\Lambda )|M({\bf P}_{12}^{c}),M({\bf P}_{34}^{c});{\bf 0%
}\rangle 
\end{eqnarray}
where 
\begin{eqnarray}
&&|M({\bf P}_{12}^{c}),M({\bf P}_{34}^{c});{\bf 0}\rangle   \nonumber \\
&=&\frac{M_{12}}{E_{12}}\int d^{3}p_{1}d^{3}p_{2}\psi ({\bf p}%
_{1}{}_{\Lambda _{12}^{-1}},{\bf p}_{2}{}_{\Lambda _{12}^{-1}})\delta ({\bf P%
}_{12}^{c}-{\bf p}_{1}-{\bf p}_{2})b^{\dagger }({\bf p}_{1})d^{\dagger }(%
{\bf p}_{2})  \nonumber \\
&&\times \frac{M_{34}}{E_{34}}\int d^{3}p_{3}d^{3}p_{4}\psi ({\bf p}%
_{3}{}_{\Lambda _{34}^{-1}},{\bf p}_{4}{}_{\Lambda _{34}^{-1}})\delta ({\bf P%
}_{34}^{c}-{\bf p}_{3}-{\bf p}_{4})b^{\dagger }({\bf p}_{3})d^{\dagger }(%
{\bf p}_{4})|0\rangle ,
\end{eqnarray}
\begin{eqnarray}
&&\langle M({\bf P}_{14}^{c}),M({\bf P}_{23}^{c});{\bf 0}|  \nonumber \\
&=&\frac{M_{14}}{E_{14}}\langle 0|\int d^{3}p_{1}^{\prime \prime
}d^{3}p_{4}^{\prime \prime }\psi ^{\ast }({\bf p}_{1}^{\prime \prime
}{}_{\Lambda _{14}^{-1}},{\bf p}_{4}^{\prime \prime }{}_{\Lambda
_{14}^{-1}})\delta ({\bf P}_{14}^{c}-{\bf p}_{1}^{\prime \prime }-{\bf p}%
_{4}^{\prime \prime })d({\bf p}_{4}^{\prime \prime })b({\bf p}_{1}^{\prime
\prime })  \nonumber \\
&&\times \frac{M_{23}}{E_{23}}\int d^{3}p_{2}^{\prime \prime
}d^{3}p_{3}^{\prime \prime }\psi ^{\ast }({\bf p}_{2}^{\prime \prime
}{}_{\Lambda _{23}^{-1}},{\bf p}_{3}^{\prime \prime }{}_{\Lambda
_{23}^{-1}})\delta ({\bf P}_{23}^{c}-{\bf p}_{2}^{\prime \prime }-{\bf p}%
_{3}^{\prime \prime })d({\bf p}_{3}^{\prime \prime })b({\bf p}_{2}^{\prime
\prime }),
\end{eqnarray}
and $E_{ij}\equiv \sqrt{{\bf P}_{ij}^{c2}+M_{ij}^{2}}$. 

Note that the integrals in the above state vectors include color summation
but not flavor. \ The above matrix element then becomes 
\begin{eqnarray}
&&\langle M({\bf P}_{14}^{c}),M({\bf P}_{23}^{c});{\bf 0}|U(\Lambda )V({\bf 
\hat{x}}_{14})U^{-1}(\Lambda )|M({\bf P}_{12}^{c}),M({\bf P}_{34}^{c});{\bf 0%
}\rangle   \nonumber \\
&=&\langle M({\bf P}_{14}^{c}),M({\bf P}_{23}^{c});{\bf 0}|\int
d^{3}p_{1}d^{3}p_{4}d^{3}p_{1}^{\prime }d^{3}p_{4}^{\prime }\delta ({\bf p}%
_{1}+{\bf p}_{4}-{\bf p}_{1}^{\prime }-{\bf p}_{4}^{\prime })\tilde{V}({\bf p%
}_{1}-{\bf p}_{4}^{\prime })  \nonumber \\
&&\times U(\Lambda )b^{\dagger }({\bf p}_{1})d^{\dagger }({\bf p}_{4})d({\bf %
p}_{4}^{\prime })b({\bf p}_{1}^{\prime })U(\Lambda )^{-1}|M({\bf P}%
_{12}^{c}),M({\bf P}_{34}^{c});{\bf 0}\rangle .
\end{eqnarray}
From our earlier arguments, the transformations defined in Eqs.\ (\ref{bbst})
and (\ref{dbst}) are independent of the total momentum of the two-body
system. Thus, using this, we obtain 
\begin{eqnarray}
&&\langle M({\bf P}_{14}^{c}),M({\bf P}_{23}^{c});{\bf 0}|U(\Lambda )V({\bf 
\hat{x}}_{14})U^{-1}(\Lambda )|M({\bf P}_{12}^{c}),M({\bf P}_{34}^{c});{\bf 0%
}\rangle   \nonumber \\
&=&\langle M({\bf P}_{14}^{c}),M({\bf P}_{23}^{c});{\bf 0}|\int
d^{3}p_{1}d^{3}p_{4}d^{3}p_{1}^{\prime }d^{3}p_{4}^{\prime }\delta ({\bf p}%
_{1}+{\bf p}_{4}-{\bf p}_{1}^{\prime }-{\bf p}_{4}^{\prime })\tilde{V}({\bf p%
}_{1}-{\bf p}_{4}^{\prime })  \nonumber \\
&&\times b^{\dagger }({\bf p}_{1\Lambda })d^{\dagger }({\bf p}_{4\Lambda })d(%
{\bf p}_{4\Lambda }^{\prime })b({\bf p}_{1\Lambda }^{\prime })|M({\bf P}%
_{12}^{c}),M({\bf P}_{34}^{c});{\bf 0}\rangle \frac{E_{14}^{2}}{M_{14}^{2}}.
\label{mtrx}
\end{eqnarray}
We can point out that this last factor, due to interaction transformations,
is not present in the nonrelativistic limit. \ This matrix element is
evaluated in Appendix B, and we find 
\begin{eqnarray}
\label{MPv14}
&&\langle M({\bf P}_{14}^{c}),M({\bf P}_{23}^{c});{\bf 0}|U(\Lambda )V({\bf 
\hat{x}}_{14})U^{-1}(\Lambda )|M({\bf P}_{12}^{c}),M({\bf P}_{34}^{c});{\bf 0
}\rangle   \nonumber \\
&&=-\delta ^{3}({\bf P}_{34}-{\bf P}_{14}+{\bf P}_{12}-{\bf P}_{23})\frac{
M_{23}}{E_{23}}\frac{E_{14}}{M_{14}}\frac{M_{12}}{E_{12}}\frac{M_{34}}{E_{34}
}  \nonumber \\
&&\times \int d^{3}p_{1}d^{3}p_{4}
\psi _{D}^{\ast }({}{\bf p}_{1\Lambda \Lambda _{14}^{-1}},
                 -{}{\bf p}_{1\Lambda \Lambda _{14}^{-1}})
\psi _{C}^{\ast }((-{\bf P}_{34}^{c}+{\bf p}_{4\Lambda
                         })_{{\bf \Lambda }_{23}^{-1}},
                 -(-{\bf P}_{34}^{c}+{\bf p}_{4\Lambda
                         })_{{\bf \Lambda }_{23}^{-1}})
\nonumber \\
&&\times
\psi _{A}({}(-{\bf P}_{23}^{c}+{\bf P}
                   _{34}^{c}-{\bf p}_{4\Lambda })_{\Lambda _{12}^{-1}},
         -{}(-{\bf P}_{23}^{c}+{\bf P}
                   _{34}^{c}-{\bf p}_{4\Lambda })_{\Lambda _{12}^{-1}}) 
\psi _{B}({}-{\bf p}_{4\Lambda }{}_{\Lambda _{34}^{-1}},
             {\bf p}_{4\Lambda }{}_{\Lambda _{34}^{-1}})
\tilde{V}({\bf p}_{1}-{\bf p}_{4}).
\end{eqnarray}
From Eqs\ (\ref{hij}), (\ref{v14}), and (\ref{MPv14}), the reaction
matrix element for the interaction $V_{14}$ is
\begin{eqnarray}
h_{fi[14]} &=&\left( \frac{M_{23}}{E_{23}}\frac{M_{12}}{E_{12}}\frac{M_{34}}{
E_{34}}\right) ^{3}\frac{M_{14}}{E_{14}}\int d^{3}p_{1}d^{3}p_{4}  
\psi_{A}( {\bf p}_A, -{\bf p}_A)
\psi_{B}( {\bf p}_B, -{\bf p}_B)
\nonumber\\
&\times&
\psi_{C}^{\ast }({\bf p}_C, -{\bf p}_C)
\psi_{D}^{\ast }({\bf p}_D, -{\bf p}_D)
\tilde{V}({\bf p}_{1}-{\bf p}_{4}),
\end{eqnarray}
where
\begin{eqnarray}
{\bf p}_A={}(-{\bf P}_{23}^{c}+{\bf P}_{34}^{c}-{\bf p}
_{4\Lambda })_{\Lambda _{12}^{-1}},
\end{eqnarray}
\begin{eqnarray}
{\bf p}_B={}-{\bf p}_{4\Lambda }{}_{\Lambda _{34}^{-1}},
\end{eqnarray}
\begin{eqnarray}
{\bf p}_C=(-{\bf P}_{34}^{c}+{\bf p}_{4\Lambda })_{{\bf \Lambda }_{23}^{-1}},
\end{eqnarray}
\begin{eqnarray}
{\bf p}_D={}{\bf p}_{1\Lambda \Lambda _{14}^{-1}}.
\end{eqnarray}
Note that the dimension of the wave function leads to an $h_{fi}$
which has the dimension of 1/mass$^{2}$ as required earlier. In the
above expression, ${}{\bf p}_{1\Lambda \Lambda _{23}^{-1}}$, for
example, is the space part of ($\Lambda _{23}^{-1}\Lambda p_{1})$ and
the four energy ratios correspond to the transformations associated
with the $(23),(14),(12),(34)$ composite particles.  The energy ratio
for the $(14)$ composite differs from the others due to the
transformation of the $(14)$ interaction term. The term here
corresponds to the $C1$ diagram in Fig. 1.

It is interesting to note that compared to the nonrelativistic case, the
overlap matrix element now involves two major differences. First, the
momentum arguments in the wave function need to be inversely boosted back to
the frame in which the composite particles are at rest, as they should be.
Second, there are factors of the type $M_{ij}/E_{ij}$ appropriate for the
composite particle in the collider frame. Both effects can lead to
substantial modification of the magnitude of the reaction cross sections.

\section{Conclusion and Summary}

We seek a relativistic formulation of the many-body problem involving
both bound states and reaction between constituents of composite
particles. As a first example, we have focused our attention on a
system of spinless particles interacting with a scalar and/or vector
interaction.

We began by examining the relativistic two-body bound state problem
and introduced Lagrangian multipliers to write down the most general
two-body Hamiltonian.  The formulation using the constraint dynamics
allowed a simple separation of the center-of-mass and the relative
motion. A two-body equation was obtained in the form of a
non-relativistic Schr\"{o}dinger equation which connects naturally to
the corresponding non-relativistic problem in the non-relativistic
limit. The two-body equation is independent of the Lagrange
multipliers. The bound state mass is related to the eigenvalue of the
non-relativistic problem by a simple algebraic equation in the case in
which one considers only relativistic kinematics. \ Further
relativistic effects show up in this algebraic relationship when the
energy dependence accompanying the scalar and vector interaction is
taken into account.

For a many-particle system, we considered pair-wise interaction
between particles. In constructing the total Hamiltonian, a good
choice of the Lagrange multiplier provides a simple way to separate
the $N$-body Hamiltonian into the unperturbed Hamiltonian and residual
interactions.  It also presents a systematic way to use the two-body
solution as basis states for multi-particle dynamics.  The study of
the dynamics involves the evaluation of the reaction matrix elements
of a general two-body interaction in terms of the wave functions of
the composite particles. 

In rearrangement reactions, because there can be many ways to divide the
total Hamiltonian, the evaluation of the reaction matrix elements should not
depend on the choice of the unperturbed Hamiltonian and basis states. With
our formulation, this ``post-prior'' equivalence can be shown explicitly,
allowing for a meaningful definition of the perturbation expansion and
treatment of the reaction dynamics.

Finally, we give an explicit formula for the reaction matrix elements
in terms of the composite wave functions. In the relativistic treatment, the
important effects include the inverse boost of the relative momentum to the
frame in which the composite particles are at rest, so as obtain the correct
wave function. Furthermore, there are factors of $M_{ij}/E_{ij}$ in the
collider frame for the composite particles. These relations will be useful
when we apply the present formulation to many problems in nuclear and
particle physics such as meson-meson scattering. \ An aim would be to see
how this approach modifies the results of the nonrelativistic formalism as
present in \cite{Bar92}.

The results we have obtained are very encouraging. We should in future
work carry out a calculation for the relativistic $I=2$ $\pi$$\pi$
scattering, to compare with the non-relativistic results of Barnes and
Swanson \cite{Bar92} and with experimental data.  We should also extend our
considerations to include the spin degree of freedom and more
complicated interactions in the constraint description.

\vfill\eject

\section*{Acknowledgments}

\vspace{-0.5cm} One of us (CYW) would like to thank Prof. Su Houng Lee for
helpful discussions and for his kind hospitality at Yonsei University where
this work was completed. This work was supported by the Department of Energy
under contract DE-AC05-00OR22725 with UT-Battelle, LLC.

\newpage \appendix

\section{Scalar Products}

\bigskip

\setcounter{section}{1}

\bigskip

Let us consider the general scalar product $\langle M({\bf P}^{\prime })|M(
{\bf P})\rangle $
\begin{eqnarray}
&\langle &M({\bf P}^{\prime })|M({\bf P})\rangle =\frac{M}{E^{\prime }}\frac{
M}{E}\int d^{3}p_{1\Lambda ^{\prime }}^{\prime }d^{3}p_{2\Lambda ^{\prime
}}^{\prime }\langle 0|d({\bf p}_{2\Lambda ^{\prime }}^{\prime })b({\bf p}
_{1\Lambda ^{\prime }}^{\prime })\psi _{M}^{\ast }({\bf p}_{1}^{\prime },
{\bf p}_{2}^{\prime })\delta ^{3}({\bf P}^{\prime }-{\bf p}_{1\Lambda
^{\prime }}^{\prime }-{\bf p}_{2\Lambda ^{\prime }}^{\prime })  \nonumber \\
&&\times \int d^{3}p_{1\Lambda }d^{3}p_{2\Lambda }\psi _{M}({\bf p}_{1},{\bf 
p}_{2})\delta ({\bf P}-{\bf p}_{1\Lambda }-{\bf p}_{2\Lambda })b^{\dagger }(
{\bf p}_{1\Lambda })d^{\dagger }({\bf p}_{2\Lambda })|0)  \nonumber \\
&=&\frac{M}{E^{\prime }}\frac{M}{E}\int d^{3}p_{1\Lambda ^{\prime }}^{\prime
}d^{3}p_{2\Lambda ^{\prime }}^{\prime }\int d^{3}p_{1\Lambda
}d^{3}p_{2\Lambda }\psi _{M}^{\ast }({\bf p}_{1}^{\prime },{\bf p}
_{2}^{\prime })\psi ({\bf p}_{1},{\bf p}_{2})\delta ^{3}({\bf P}^{\prime }-
{\bf P})\delta ^{3}({\bf P}-{\bf p}_{1\Lambda }-{\bf p}_{2\Lambda }) 
\nonumber \\
&&\times \delta ^{3}({\bf p}_{1\Lambda }-{\bf p}_{1\Lambda ^{\prime
}}^{\prime })\delta ^{3}({\bf p}_{2\Lambda }-{\bf p}_{2\Lambda ^{\prime
}}^{\prime }),
\end{eqnarray}
where $E=\sqrt{{\bf P}^{2}+M^{2}},$ $E^{\prime }=\sqrt{{\bf P}^{\prime
2}+M^{2}}.$

Now the total momentum delta function makes $\Lambda =\Lambda ^{\prime }$,
which in turn implies that the two delta functions that come from the
creation and annihilation operator force the arguments of the two wave
functions to be the same. \ Thus 
\begin{eqnarray}
&\langle&M({\bf P}^{\prime })|M({\bf P})\rangle=\delta ^{3}({\bf P}^{\prime
}-{\bf P})\frac{M^{2}}{E^{2}}\int d^{3}p_{1\Lambda }d^{3}p_{2\Lambda }\psi
_{M}^{\ast }({\bf p}_{1},{\bf p}_{2})\psi _{M}({\bf p}_{1},{\bf p}%
_{2})\delta ^{3}({\bf P}-{\bf p}_{1\Lambda }-{\bf p}_{2\Lambda })  \nonumber
\\
&=&\delta ^{3}({\bf P}^{\prime }-{\bf P})\frac{M^{2}}{E^{2}}\int
d^{3}p_{1}d^{3}p_{2}\psi _{M}^{\ast }({\bf p}_{1\Lambda ^{-1}},{\bf p}%
_{2\Lambda ^{-1}})\psi _{M}({\bf p}_{1\Lambda ^{-1}},{\bf p}_{2\Lambda
^{-1}})\delta ^{3}({\bf P}-{\bf p}_{1}-{\bf p}_{2}).
\end{eqnarray}
On the right-hand side let 
\begin{eqnarray}
{\bf P} &=&{\bf p}_{1}+{\bf p}_{2},  \nonumber \\
{\bf p} &=&\frac{\varepsilon _{2}}{M}{\bf p}_{1}-\frac{\varepsilon _{1}}{M}%
{\bf p}_{2},
\end{eqnarray}
so that the integral becomes 
\begin{eqnarray}
&&\int d^{3}Pd^{3}p\psi _{M}^{\ast }((\frac{\varepsilon _{1}}{M}{\bf P}+{\bf %
p})_{\Lambda ^{-1}},((\frac{\varepsilon _{2}}{M}{\bf P}-{\bf p}))_{\Lambda
^{-1}})\psi _{M}((\frac{\varepsilon _{1}}{M}{\bf P}+{\bf p})_{\Lambda
^{-1}},((\frac{\varepsilon _{2}}{M}{\bf P}-{\bf p}))_{\Lambda ^{-1}})\delta (%
{\bf P}-{\bf P})  \nonumber \\
&=&\int d^{3}p\psi _{M}^{\ast }((\frac{\varepsilon _{1}}{M}{\bf P}+{\bf p}%
)_{\Lambda ^{-1}},((\frac{\varepsilon _{2}}{M}{\bf P}-{\bf p}))_{\Lambda
^{-1}})\psi _{M}((\frac{\varepsilon _{1}}{M}{\bf P}+{\bf p})_{\Lambda
^{-1}},((\frac{\varepsilon _{2}}{M}{\bf P}-{\bf p}^{\prime }))_{\Lambda
^{-1}})  \nonumber \\
&=&\int d^{3}p\psi _{M}^{\ast }(({\bf 0}+{\bf p}^{\prime })_{\Lambda
^{-1}},(({\bf 0}-{\bf p}))_{\Lambda ^{-1}})\psi _{M}(({\bf 0}+{\bf p}%
)_{\Lambda ^{-1}},(({\bf 0}-{\bf p}))_{\Lambda ^{-1}})  \nonumber \\
&\equiv &\int d^{3}p|\psi _{M}({\bf p}_{\Lambda ^{-1}})|^{2}=\int
d^{4}p\delta (p^{0})|\psi _{M}(\Lambda ^{-1}p)|^{2}.
\end{eqnarray}
But $p^{0}=p\cdot \Lambda ^{-1}P/M$. \ So taking $p=\Lambda \bar{p}$ and
using $d^{4}p^{\prime }=d^{4}\bar{p}$, we have the following manifestly
covariant scalar product 
\begin{equation}
\int d^{4}p\delta (p^{0})|\psi _{M}(\Lambda ^{-1}p)|^{2}=M\int d^{4}\bar{p}
\delta (\bar{p}\cdot P^{\prime })|\psi _{M}(\bar{p})|^{2}.
\end{equation}

\section{\protect\bigskip Evaluation of the Reaction Matrix Element 
}

In this Appendix we evaluate the matrix element of Eq.\ (\ref{mtrx}),
\begin{eqnarray}
&&\langle M({\bf P}_{14}^{c}),M({\bf P}_{23}^{c});{\bf 0}|U(\Lambda )V({\bf 
\hat{x}}_{14})U^{-1}(\Lambda )|M({\bf P}_{12}^{c}),M({\bf P}_{34}^{c});{\bf 0
}\rangle   \nonumber \\
&=&\langle M({\bf P}_{14}^{c}),M({\bf P}_{23}^{c});{\bf 0}|\int
d^{3}p_{1}^{\prime \prime \prime }d^{3}p_{4}^{\prime \prime \prime
}d^{3}p_{1}^{\prime }d^{3}p_{4}^{\prime }\delta ({\bf p}_{1}^{\prime \prime
\prime }+{\bf p}_{4}^{\prime \prime \prime }-{\bf p}_{1}^{\prime }-{\bf p}%
_{4}^{\prime })\tilde{V}({\bf p}_{1}^{\prime \prime \prime }-{\bf p}%
_{4}^{\prime })  \nonumber \\
&&\times U(\Lambda )b^{\dagger }({\bf p}_{1}^{\prime \prime \prime
})d^{\dagger }({\bf p}_{4}^{\prime \prime \prime })d({\bf p}_{4}^{\prime })b(%
{\bf p}_{1}^{\prime })U(\Lambda )^{-1}|M({\bf P}_{12}^{c}),M({\bf P}%
_{34}^{c});{\bf 0}\rangle   \nonumber \\
&=&\langle M({\bf P}_{14}^{c}),M({\bf P}_{23}^{c});{\bf 0}|\int
d^{3}p_{1}^{\prime \prime \prime }d^{3}p_{4}^{\prime \prime \prime
}d^{3}p_{1}^{\prime }d^{3}p_{4}^{\prime }\delta ({\bf p}_{1}^{\prime \prime
\prime }+{\bf p}_{4}^{\prime \prime \prime }-{\bf p}_{1}^{\prime }-{\bf p}%
_{4}^{\prime })\tilde{V}({\bf p}_{1}^{\prime \prime \prime }-{\bf p}%
_{4}^{\prime })  \nonumber \\
&&b^{\dagger }({\bf p}_{1\Lambda }^{\prime \prime \prime })d^{\dagger }({\bf %
p}_{4\Lambda }^{\prime \prime \prime })d({\bf p}_{4\Lambda }^{\prime })b(%
{\bf p}_{1\Lambda }^{\prime })|M({\bf P}_{12}^{c}),M({\bf P}_{34}^{c});{\bf 0%
}\rangle \frac{{\bf P}_{14}^{c\prime 2}+M_{14}^{2}}{M_{14}^{2}},
\end{eqnarray}
in which the integrals include sums over flavor and color. This
requires us to compute
\begin{eqnarray}
&&\langle 0|d({\bf p}_{4}^{\prime \prime })b({\bf p}_{1}^{\prime \prime })d(
{\bf p}_{3}^{\prime \prime })b({\bf p}_{2}^{\prime \prime })b^{\dagger }(
{\bf p}_{1\Lambda }^{\prime \prime \prime })d^{\dagger }({\bf p}_{4\Lambda
}^{\prime \prime \prime })d({\bf p}_{4\Lambda }^{\prime })b({\bf p}
_{1\Lambda }^{\prime })b^{\dagger }({\bf p}_{1})d^{\dagger }({\bf p}
_{2})b^{\dagger }({\bf p}_{3})d^{\dagger }({\bf p}_{4})|0\rangle   \nonumber
\\
&=&\langle 0|[\delta ({\bf p}_{4\Lambda }^{\prime \prime \prime }-{\bf p}
_{3}^{\prime \prime })\delta ({\bf p}_{1\Lambda }^{\prime \prime \prime }-
{\bf p}_{2}^{\prime \prime })d({\bf p}_{4}^{\prime \prime })b({\bf p}
_{1}^{\prime \prime })-\delta ({\bf p}_{4\Lambda }^{\prime \prime \prime }-
{\bf p}_{4}^{\prime \prime })\delta ({\bf p}_{1\Lambda }^{\prime \prime
\prime }-{\bf p}_{2}^{\prime \prime })d({\bf p}_{3}^{\prime \prime })b({\bf p
}_{1}^{\prime \prime })  \nonumber \\
&&-\delta ({\bf p}_{4\Lambda }^{\prime \prime \prime }-{\bf p}_{3}^{\prime
\prime })\delta ({\bf p}_{1\Lambda }^{\prime \prime \prime }-{\bf p}
_{1}^{\prime \prime })d({\bf p}_{4}^{\prime \prime })b({\bf p}_{2}^{\prime
\prime })+\delta ({\bf p}_{4\Lambda }^{\prime \prime \prime }-{\bf p}
_{4}^{\prime \prime })\delta ({\bf p}_{1\Lambda }^{\prime \prime \prime }-
{\bf p}_{1}^{\prime \prime })d({\bf p}_{3}^{\prime \prime })b({\bf p}
_{2}^{\prime \prime })]  \nonumber \\
&&\times \lbrack \delta ({\bf p}_{4\Lambda }^{\prime }-{\bf p}_{2})\delta (
{\bf p}_{1\Lambda }^{\prime }-{\bf p}_{1})b^{\dagger }({\bf p}
_{3})d^{\dagger }({\bf p}_{4})-\delta ({\bf p}_{4\Lambda }^{\prime }-{\bf p}
_{4})\delta ({\bf p}_{1\Lambda }^{\prime }-{\bf p}_{1})b^{\dagger }({\bf p}
_{3})d^{\dagger }({\bf p}_{2})  \nonumber \\
&&-\delta ({\bf p}_{4\Lambda }^{\prime }-{\bf p}_{2})\delta ({\bf p}
_{1\Lambda }^{\prime }-{\bf p}_{3})b^{\dagger }({\bf p}_{1})d^{\dagger }(
{\bf p}_{4})+\delta ({\bf p}_{4\Lambda }^{\prime }-{\bf p}_{4})\delta ({\bf p
}_{1\Lambda }^{\prime }-{\bf p}_{3})b^{\dagger }({\bf p}_{1})d^{\dagger }(
{\bf p}_{2})]|0 \rangle
\end{eqnarray}
in which the delta functions include flavor and color indices. \ If we
assume that flavors for 1 and 2 are distinct from those of 3 and 4, then of
the sixteen terms above the only one that survives is 
\begin{eqnarray}
\langle 0|\delta ({\bf p}_{4\Lambda }^{\prime \prime \prime }-{\bf p}%
_{4}^{\prime \prime })\delta ({\bf p}_{1\Lambda }^{\prime \prime \prime }-%
{\bf p}_{1}^{\prime \prime })\delta ({\bf p}_{4\Lambda }^{\prime }-{\bf p}%
_{4})\delta ({\bf p}_{1\Lambda }^{\prime }-{\bf p}_{1})\delta ({\bf p}%
_{3}^{\prime \prime }-{\bf p}_{3})\delta ({\bf p}_{2}^{\prime \prime }-{\bf p%
}_{2})|0).
\end{eqnarray}
Thus (using the notation $E_{ij}=\sqrt{{\bf
P}_{ij}^{c2}+M_{ij}^{2}}$), we obtain
\bigskip 
\begin{eqnarray}
&&\langle M({\bf P}_{14}^{c}),M({\bf P}_{23}^{c});{\bf 0}|U(\Lambda )V({\bf 
\hat{x}}_{14})U^{-1}(\Lambda )|M({\bf P}_{12}^{c}),M({\bf P}_{34}^{c});{\bf 0%
}\rangle   \nonumber \\
= &&-\int d^{3}p_{4}^{\prime \prime \prime }d^{3}p_{1}^{\prime \prime \prime
}d^{3}p_{4}^{\prime }d^{3}p_{1}^{\prime }d^{3}p_{3}^{\prime \prime
}d^{3}p_{2}^{\prime \prime }\psi _{D}^{\ast }({}{\bf p}_{1\Lambda \Lambda
_{14}^{-1}}^{\prime \prime \prime },{}{\bf p}_{4\Lambda \Lambda
_{14}^{-1}}^{\prime \prime \prime })\delta ({\bf P}_{14}^{c}-{}{\bf p}%
_{1\Lambda }^{\prime \prime \prime }-{}{\bf p}_{4\Lambda }^{\prime \prime
\prime })\psi _{C}^{\ast }({}{\bf p}_{2\Lambda _{23}^{-1}}^{\prime \prime },%
{\bf p}_{3}^{\prime \prime }{}_{\Lambda _{23}^{-1}})  \nonumber \\
&&\times \delta ({\bf P}_{23}^{c}-{\bf p}_{2}^{\prime \prime }-{\bf p}%
_{3}^{\prime \prime })\psi _{A}({}{\bf p}_{1\Lambda \Lambda
_{12}^{-1}}^{\prime },{\bf p}_{2}^{\prime \prime }{}_{\Lambda
_{12}^{-1}})\delta ({\bf P}_{12}^{c}-{\bf p}_{2}^{\prime \prime }-{\bf p}%
_{1\Lambda }^{\prime })\psi _{B}({\bf p}_{3}^{\prime \prime }{}_{\Lambda
_{34}^{-1}},{\bf p}_{4\Lambda }^{\prime }{}_{\Lambda _{34}^{-1}})\delta (%
{\bf P}_{34}^{c}-{\bf p}_{4\Lambda }^{\prime }-{\bf p}_{3}^{\prime \prime })
\nonumber \\
&&\times \delta ({\bf p}_{1}^{\prime \prime \prime }+{\bf p}_{4}^{\prime
\prime \prime }-{\bf p}_{1}^{\prime }-{\bf p}_{4}^{\prime })\tilde{V}({\bf p}%
_{1}^{\prime \prime \prime }-{\bf p}_{4}^{\prime })\frac{M_{23}}{E_{23}}%
\frac{E_{14}}{M_{14}}\frac{M_{12}}{E_{12}}\frac{M_{34}}{E_{34}}.
\end{eqnarray}
We perform four of the remaining six volume integrals of the first set of
integrals. In particular, we perform integrations over $d^{3}p_{4}^{\prime
\prime \prime }d^{3}p_{1}^{\prime }d^{3}p_{3}^{\prime \prime
}d^{3}p_{2}^{\prime \prime }$. \ We use the first delta function ${\bf p}%
_{4\Lambda }^{\prime \prime \prime }={\bf P}_{14}^{c}-{\bf p}_{1\Lambda
}^{\prime \prime \prime }$ , the second gives ${\bf p}_{2}^{\prime \prime }=%
{\bf P}_{23}^{c}-{\bf p}_{3}^{\prime \prime }$, the third gives ${\bf p}%
_{1\Lambda }^{\prime }={\bf P}_{12}^{c}-{\bf p}_{2}^{\prime \prime }$ \ and
the fourth gives ${\bf p}_{3}^{\prime \prime }={\bf P}_{34}^{c}-{\bf p}%
_{4\Lambda }^{\prime }.$ The argument of the remaining delta function is
then 
\begin{equation}
{\bf p}_{1}^{\prime \prime \prime }+{\bf p}_{4}^{\prime \prime \prime }-{\bf %
p}_{1}^{\prime }-{\bf p}_{4}^{\prime }={\bf P}_{14\Lambda ^{-1}}^{c}-{\bf P}%
_{12\Lambda ^{-1}}^{c}+{\bf P}_{34\Lambda ^{-1}}^{c}-{\bf P}_{23\Lambda
^{-1}}^{c}={\bf P}_{34}-{\bf P}_{14}+{\bf P}_{12}-{\bf P}_{23},
\end{equation}
which corresponds to overall momentum conservation in the frame in which we
evaluate the matrix element (the ${\bf P}_{14}=0$ frame). 
Then, the reaction matrix element becomes
\begin{eqnarray}
&&\langle M({\bf P}_{14}^{c}),M({\bf P}_{23}^{c});{\bf 0}|U(\Lambda )V({\bf 
\hat{x}}_{14})U^{-1}(\Lambda )|M({\bf P}_{12}^{c}),M({\bf P}_{34}^{c});{\bf 0%
}\rangle   \nonumber \\
= &&-\delta ^{3}({\bf P}_{34}-{\bf P}_{14}+{\bf P}_{12}-{\bf P}_{23})\int
d^{3}p_{1}^{\prime \prime \prime }d^{3}p_{4}^{\prime }\psi _{D}^{\ast }({}%
{\bf p}_{1\Lambda \Lambda _{14}^{-1}}^{\prime \prime \prime },{}({\bf P}%
_{14}^{c}-{\bf p}_{1\Lambda }^{\prime \prime \prime })_{\Lambda _{14}^{-1}})
\nonumber \\
&&\times \psi _{C}^{\ast }({\bf P}_{23}^{c}{}-{\bf P}_{34}^{c}+{\bf p}%
_{4\Lambda }^{\prime })_{{\bf \Lambda }_{23}^{-1}},({\bf P}_{34}^{c}-{\bf p}%
_{4\Lambda }^{\prime })_{\Lambda _{23}^{-1}})\psi _{A}({}({\bf P}_{12}^{c}-%
{\bf P}_{23}^{c}+{\bf P}_{34}^{c}-{\bf p}_{4\Lambda }^{\prime })_{\Lambda
_{12}^{-1}},({\bf P}_{23}^{c}{}-{\bf P}_{34}^{c}+{\bf p}_{4\Lambda }^{\prime
})_{\Lambda _{12}^{-1}})  \nonumber \\
&&\times \psi _{B}({}({\bf P}_{34}^{c}-{\bf p}_{4\Lambda }^{\prime
})_{\Lambda _{34}^{-1}},{\bf p}_{4\Lambda }^{\prime }{}_{\Lambda _{34}^{-1}})%
\tilde{V}({\bf p}_{1}^{\prime \prime \prime }-{\bf p}_{4}^{\prime })\frac{%
M_{23}}{E_{23}}\frac{E_{14}}{M_{14}}\frac{M_{12}}{E_{12}}\frac{M_{34}}{E_{34}%
}.
\end{eqnarray}
But ${\bf P}_{14}{}_{\Lambda _{14}^{-1}}=0={\bf P}_{12}{}_{\Lambda
_{12}^{-1}}$ $={\bf P}_{34}{}_{\Lambda _{34}^{-1}}$ $={\bf P}%
_{23}{}_{\Lambda _{23}^{-1}}$ so that 
\begin{eqnarray}
&&\langle M({\bf P}_{14}^{c}),M({\bf P}_{23}^{c});{\bf 0}|U(\Lambda )V({\bf 
\hat{x}}_{14})U^{-1}(\Lambda )|M({\bf P}_{12}^{c}),M({\bf P}_{34}^{c});{\bf 0%
}\rangle   \nonumber \\
&&=-\delta ^{3}({\bf P}_{34}-{\bf P}_{14}+{\bf P}_{12}-{\bf P}_{23})\int
d^{3}p_{1}^{\prime \prime \prime }d^{3}p_{4}^{\prime }\psi _{D}^{\ast }({}%
{\bf p}_{1\Lambda \Lambda _{14}^{-1}}^{\prime \prime \prime },{}-{\bf p}%
_{1\Lambda }^{\prime \prime \prime }{}_{\Lambda _{14}^{-1}})\psi _{C}^{\ast
}((-{\bf P}_{34}^{c}+{\bf p}_{4\Lambda }^{\prime })_{{\bf \Lambda }%
_{23}^{-1}},({\bf P}_{34}^{c}-{\bf p}_{4\Lambda }^{\prime })_{\Lambda
_{23}^{-1}})  \nonumber \\
&&\times \tilde{V}({\bf p}_{1}^{\prime \prime \prime }-{\bf p}_{4}^{\prime
})\psi _{A}({}(-{\bf P}_{23}^{c}+{\bf P}_{34}^{c}-{\bf p}_{4\Lambda
}^{\prime })_{\Lambda _{12}^{-1}},({\bf P}_{23}^{c}{}-{\bf P}_{34}^{c}+{\bf p%
}_{4\Lambda }^{\prime })_{\Lambda _{12}^{-1}})\psi _{B}({}-{\bf p}_{4\Lambda
}^{\prime }{}_{\Lambda _{34}^{-1}},{\bf p}_{4\Lambda }^{\prime }{}_{\Lambda
_{34}^{-1}})\frac{M_{23}}{E_{23}}\frac{E_{14}}{M_{14}}\frac{M_{12}}{E_{12}}%
\frac{M_{34}}{E_{34}},
\end{eqnarray}
which corresponds to the indicated amplitude in which a gluon is
exchanged between particle $1$ and particle ${4}.$ This
produces the form in the text (where we dropped the primes on the two
integration variables).


\begin{references}
\bibitem{time}  This is the time component of the relativistic center of
mass position operator $X^{\mu }$ satisfying $[X^{\mu },P^{\nu }]=i\eta ^{\mu
\nu }$ and whose classical form is $X^{\mu }=(\varepsilon _{1}x_{1}^{\mu
}+\varepsilon _{2}x_{2}^{\mu })/M+(m_{1}^{2}-m_{2}^{2})x\cdot PP^{\mu
}/(P^{2})^{2}$ \ (see \cite{Cra84} p 2590). \ This commutator leads to $%
\langle t|P^{0}\rangle =\exp (-iP^{0}t)$.
\end{references}
\end{document}